**Interplay between topology and correlations in the second moiré band of twisted bilayer $MoTe_2$**

Fan Xu[1†], Xumin Chang[1†], Jiayong Xiao[1†], Yixin Zhang[2,3†], Feng Liu[1], Zheng Sun[1], Ning Mao[2], Nikolai Peshcherenko[2], Jiayi Li[1], Kenji Watanabe[4], Takashi Taniguchi[5], Bingbing Tong[6], Li Lu[6,7], Jinfeng Jia[1,7,8], Dong Qian[1], Zhiwen Shi[1], Yang Zhang[9,10*], Xiaoxue Liu[1,7,8*], Shengwei Jiang[1*], and Tingxin Li[1,7*]

[1]Key Laboratory of Artificial Structures and Quantum Control (Ministry of Education), Tsung-Dao Lee Institute, School of Physics and Astronomy, Shanghai Jiao Tong University, Shanghai, China

[2]Max Planck Institute for Chemical Physics of Solids, Dresden, Germany

[3]School of Physics, Peking University, Beijing, China

[4]Research Center for Electronic and Optical Materials, National Institute for Materials Science, 1-1 Namiki, Tsukuba, Japan

[5]Research Center for Materials Nanoarchitectonics, National Institute for Materials Science, 1-1 Namiki, Tsukuba, Japan

[6]Beijing National Laboratory for Condensed Matter Physics and Institute of Physics, Chinese Academy of Sciences, Beijing, China

[7]Hefei National Laboratory, Hefei, China

[8]Shanghai Research Center for Quantum Sciences, Shanghai, China

[9]Department of Physics and Astronomy, University of Tennessee, Knoxville, TN, USA

[10]Min H. Kao Department of Electrical Engineering and Computer Science, University of Tennessee, Knoxville, Tennessee, USA

[†]These authors contribute equally to this work.

[*]Emails: yangzhang@utk.edu, xxliu90@sjtu.edu.cn, swjiang@sjtu.edu.cn, txli89@sjtu.edu.cn,

**Abstract**

**Topological flat bands formed in two-dimensional lattice systems offer unique opportunity to study the fractional phases of matter in the absence of an external magnetic field. Celebrated examples include fractional quantum anomalous Hall (FQAH) effects[1-5] and fractional topological insulators[6]. Recently, FQAH effects have been experimentally realized in both the twisted bilayer $MoTe_2$ ($tMoTe_2$) system[7-10] and the rhombohedral stacked multilayer graphene/hBN moiré**

**system[11,12]. To date, experimental studies mainly focus on the first moiré flat band, except a very recent work[13] that studied novel transport properties in higher moiré bands of a 2.1° $tMoTe_2$ device. Here, we present the systematical transport study of approximately 3° $tMoTe_2$ devices, especially for the second moiré band. At $\nu$ = -2 and -4, time-reversal-symmetric single and double quantum spin Hall states formed, consistent with the previous observation in 2.1° $tMoTe_2$ device[13]. On the other hand, we observed ferromagnetism in the second moiré band, and a Chern insulator state driven by out-of-plane magnetic fields at $\nu$ = -3. At $\nu \approx$ -2.2 to -2.7, finite temperature resistivity minimum with $1/T$ scaling at low temperatures, and large out-of-plane negative magnetoresistance have been observed. Applying out-of-plane electric field can induce quantum phase transitions at both integer and fractional filling factors. Our studies pave the way for realizing tunable topological states and other unexpected magnetic phases beyond the first moiré flat band based on twisted $MoTe_2$ platform.**

## Introduction

Semiconducting transition metal dichalcogenide (TMDc) moiré superlattices have emerged as a highly tunable platform for exploring correlated and topological quantum phases of matter. In particular, highly tunable topological flat bands have been theoretically proposed[14-18], and experimentally realized in both TMDc moiré heterobilayers[19-21] and twisted TMDc homobilayers[7-10,13,22-26]. Recently, FQAH effect has been experimentally observed[7-10] in $tMoTe_2$ systems, attracting intense attentions[27-58].

Previous experimental studies of $tMoTe_2$ have mainly focused on two different twist angle ranges, each regime exhibiting distinct yet remarkable topological quantum states. For relatively large twist angles[7-10] (about 3.5° to 4°), within the first moiré flat band, integer quantum anomalous Hall (IQAH) effect has been observed at $\nu$ = -1, and FQAH states have been observed at $\nu$ = -2/3 and -3/5. Theoretically, the global phase diagram ($\nu$ = -1 IQAH, $\nu$ = -2/3 FQAH state, $\nu$ = -1/2 composite Fermi liquid and $\nu$ = -1/3 charge density wave) at large twist angles has been well studied from electronic structure theory[34-36, 40, 45], Hartree-Fock calculation[30, 32, 37, 41-44, 47], density matrix renormalization group (DMRG)[30], and in single band[29, 30, 31, 33, 38, 46] and band-mixing[39, 44, 48] exact diagonalizations. While for small twist angles[13] (2.1°-2.2°), besides the observation of the IQAH effect at $\nu$ = -1, a series of integer quantum spin Hall (IQSH) insulators have been identified at $\nu$ = -2, -4, and -6. More importantly, evidence of a fractional quantum spin Hall (FQSH) insulator at $\nu$ = -3 has also been observed. The higher moiré bands open the door for engineering bosonic quantum spin Hall effect from the coupling of multiple quantum spin Hall edge states, and non-Abelian phases[49-58] such as Moore-Read states in analogy to even denominator fractional quantum Hall states in the second Landau level. Here, we mainly report the transport studies of $tMoTe_2$ devices with intermediary twist angles (around 3°) compared to previous works. We observed a series of quantum phases arising from the interplay between topology and strong correlations in the second moiré flat band.

## Phase diagram of approximately 3° $tMoTe_2$

The schematic design of our $tMoTe_2$ devices is shown in Fig. 1a. The device has a standard Hall bar geometry with three gates, where the $Si/SiO_2$ gate is used to heavily dope $tMoTe_2$ to form Ohmic contacts with $TaSe_2$, while the top and bottom graphite gates allow for independent control of the moiré filling factor $\nu$ and the out-of-plane electric displacement field $D$ (Methods). Figure 1b shows the calculated band structure of a 3.15° $tMoTe_2$ (Methods), where the first two moiré bands are well isolated with Chern number $C = +1$. The bandwidth of the first two moiré bands is about 8 meV and 15 meV, respectively. Figure 1c and 1d show the longitudinal resistivity $\rho_{xx}$ and Hall resistivity $\rho_{xy}$, respectively, as a function of $\nu$ and $D$ of a 3.15° $tMoTe_2$ device (moiré density $n_M \approx 2.7 \times 10^{12}\ \mathrm{cm}^{-2}$, denoted as device A), where the twist angle is calibrated based on the quantum oscillations emerging under large out-of-plane magnetic field $B_\perp$ (Extended Data Fig. 1). For $\nu$ = 0 to -1.25, the observed features closely resemble previously reported phase diagram[9,10], including the IQAH state at $\nu$ = -1, the FQAH state at $\nu$ = -2/3, and trivial correlated insulating states under certain $D$-fields at these fillings. At higher $\nu$, a number of new features emerged in both $\rho_{xx}$ and $\rho_{xy}$, which are the main focus of this paper. Similar results have been observed in device B with a twist angle about 3.0° (Extended Data Fig. 2).

## Magnetism in the second moiré band

Figure 2a displays the symmetrized $\rho_{xx}$ and antisymmetrized $\rho_{xy}$ at $B_\perp = \pm 0.1$ T and $\pm 0.3$ T at $D \approx 0$ of device A (3.15°). At $\nu$ = -1, we observed quantized $\rho_{xy}$ plateau of $h/e^2$ and vanishing $\rho_{xx}$, consistent with previously reported IQAH effect in $tMoTe_2$ system [9,10,13]. At $\nu$ = -2/3, a weaker Hall plateau around $3h/2e^2$ (±2.5% at $B_\perp = \pm 0.3$ T and $T$ = 1.2 K) was observed, accompanied by a minimum in $\rho_{xx}$ of approximately 6.5 kΩ, consistent with previous reported[9,10] FQAH effect in relatively large-angle (3.7°-3.9°) $tMoTe_2$. The $\rho_{xy}$ plateau of the FQAH state gets stronger at lower temperature and higher $B_\perp$ (inset of Fig. 2a). Overall, ferromagnetism in the first moiré band spans approximately from $\nu \approx$ -1.5 to -0.4.

Remarkably, anomalous Hall (AH) signals have also been observed in the second moiré band. Figure 2b shows the $\rho_{xy}$ versus $B_{\perp}$ at various $\nu$ within the second moiré band of device A (3.15°). Enhanced Hall signals at small $B_{\perp}$ have been observed from $\nu \approx -2.4$ to -3.2, whereas magnetic hysteresis and zero-field AH can only be identified from $\nu \approx -2.7$ to -3.2. The AH signal in the second moiré band is opposite to that in the first moiré band, which can be understood from the non-uniform Berry curvature distribution in the dispersive second moiré band (Methods). We found that the magnetism of the second moiré band still persists in devices with larger twisted angle (see Extended Data Fig. 3 for results of a 3.7° $tMoTe_2$ device, denoted as device C). The Curie temperature $T_c$ of the magnetism in the second moiré band is about 4-5 K (Extended Data Fig. 3 and 4), which is lower than that of the first moiré band ($T_c$ > 10 K at $\nu$ = 1) [7-10]. The emergence of magnetism in the second moiré band is also evident by the magnetic circular dichroism (MCD) measurements (Extended Data Fig. 3 and 4). Similar to the first moiré band, applied $D$-fields could quench the ferromagnetism in the second moiré band. Theoretically, the second moiré band of 3.15° $tMoTe_2$ is also topologically nontrivial, and has a direct gap only around $D \approx 0$ V/nm, indicating the observed ferromagnetism in the second moiré band also has a topological origin (Methods). On the other hand, in current devices, we didn't observe any clear signatures (namely the quantized $\rho_{xy}$ and/or local $\rho_{xx}$ dip) for the formation of FQAH states in the second moiré band.

## IQSH phases at $\nu$ = -2, -4

In the recent study[13] of 2.1° $tMoTe_2$, quantized two-terminal resistance plateaus of $h/2e^2$, $h/3e^2$, $h/4e^2$, and $h/6e^2$ have been observed under zero magnetic field at $\nu$ = -2, -3, -4, and -6, respectively, where the Hall resistance $R_{xy}$ at these fillings stays around zero. These observations indicate the formation of IQSH insulators at $\nu$ = -2, -4, -6, and a FQSH insulator at $\nu$ = -3. For quantum spin Hall insulators formed in TMDc moiré superlattices[13,19,20,23], the helical edge states are supposed to possess Ising-like (out-of-plane) spins, thus they have been protected by the spin-$S_z$ conservation symmetry rather than the time-reversal symmetry (TRS) in typical two-dimensional topological insulators[59]. Consequently, the magnetoresistance of helical edges is highly anisotropic, as reported in previous studies of AB-stacked $WSe_2/MoTe_2$ [19], twisted bilayer $WSe_2$ [23], and $tMoTe_2$ [13] moiré systems. In $tMoTe_2$ devices with twist angle range from 3.0° to 3.7°, at both $\nu$ = -2 and -4, we have also observed a strong magnetoresistance (up to ~400%) under a small in-plane magnetic field $B_{\parallel}$, together with a weak magnetoresistance under $B_{\perp}$ (Extended Data Fig. 5). Figure 3a illustrates the in-plane longitudinal magnetoresistance $R_{xx}(B_{\parallel}=0.3\ \mathrm{T})/R_{xx}(B=0)$ versus $\nu$ and $D$ in device A (3.15°). Moreover, by studying the dependence of magnetoresistance on the direction of the applied magnetic field in device B (3.0°) and device C (3.7°) (Extended Data Fig.

6), we found the magnetoresistance at $\nu$ = -2 and -4 is indeed maximized (minimized) when the magnetic field is applied parallel (perpendicular) to the device. These observations are consistent with the expectation of spin-$S_z$ conserved IQSH insulators, where the helical edge transport is immune to applied magnetic fields along the spin quantization axis[13,19,23,60]. The presence of helical edge states is also supported by the non-local transport, where the ratio between nonlocal resistance and local resistance is significantly enhanced at both $\nu$ = -2 and -4, as shown in Fig. 3b and Extended Data Fig. 7. However, we didn't observe the quantized plateaus of longitudinal resistance $R_{xx}$ associated with the ballistic transport of the helical edge states, presumably due to the moiré inhomogeneity and/or other disorder effects.

Applied $D$-field could drive the IQSH phase to a trivial insulator phase at $\nu$ = -2. The transport gap of the IQSH phase and the trivial insulator phase is estimated to be about 1 meV and 2.5 meV, respectively (Extended Data Fig. 5). These are consistent with the small single-particle energy gaps between the first two moiré bands (Extended Data Fig. 8). Within the Hartree-Fock approximation, we find that $\nu$ = -2 undergoes a transition from IQSH insulator to antiferromagnetic insulator in the strong interaction regime, and from IQSH to nonmagnetic trivial insulator in the weak interaction regime (Extended Data Fig. 8,9).

### $B_\perp$ and $D$-field induced quantum phase transitions at $\nu$ = -3

At $\nu$ = -3, we found the transport behavior in approximately 3° $tMoTe_2$ devices is qualitatively different from that in the 2.1° $tMoTe_2$ device[13]. Firstly, in contrast to the time-reversal-symmetric phase with zero $R_{xy}$ reported in Ref. 13, clear ferromagnetism featured by anomalous Hall signals has been observed at $\nu$ = -3 in approximately 3° $tMoTe_2$ (Fig. 2). Secondly, the temperature dependence of $\rho_{xx}$ at $\nu$ = -3 and $D \approx 0$ in approximately 3° $tMoTe_2$ devices exhibit metallic behavior instead of insulating behavior (Fig. 3c for device A and Extended Data Fig. 10 for device B), although a weak $\rho_{xx}$ peak can be identified around $\nu$ = -3 (Fig. 2a). Overall, in approximately 3° $tMoTe_2$ devices, transport results at $\nu$ = -3 resembles an AH metal phase. Interestingly, the application of $B_\perp$ drives the AH metal into a Chern insulating state with Chern number $C \approx$ -1, which is determined from the $\rho_{xy}$ values and the Streda formula $n_M \frac{d\nu}{dB_\perp} = C\frac{e}{h}$ (Fig. 3d for device A and Extended Data Fig. 10 for device B). We noticed that the Chern number of the $\nu$ = -3 Chern state is opposite to the $C$ = 1 IQAH state at $\nu$ = -1. At $B_\perp$= 0, applied $D$-field could drive the AH metal phase to a resistive phase, and further transits into a nonmagnetic metal phase.

We performed band structure and density of state (DOS) calculations under different $D$-fields, and found a pronounced van Hove singularity (VHS) peak at $\nu$ = -3, which is smeared out by applied $D$-field, as shown in Extended Data Fig 8. Therefore, the AH metal behavior and the local $\rho_{xx}$ peak are likely attributed to the instability from DOS divergence at $\nu$ = -3. For the nature of TRS breaking Chern states, we performed Hartree-Fock calculations at zero magnetic field and also with Zeeman energy of 2 meV (Methods and Extended Data Fig. 9). At filling factor $\nu$ = -3 without magnetic field, we found a weakly polarized metal state with finite anomalous Hall conductivity. As interactions weaken, the ground state becomes valley polarized with a quantized Hall response, where the first moiré band at K' valley and the first two moiré bands at K valley have been occupied. The state exhibits a nearly zero direct gap between an occupied band in K valley and an unoccupied band in K' valley. When a small Zeeman energy is applied, the gap is enlarged. The Chern number of the top two moiré bands at K valley changes from (1,1) to (1,-1) due to Hatree-Fock renormalizations, resulting in a $C$ = -1 Chern insulating state, consistent with the experimental observations. Overall, at $\nu$ = -3 and $B_{\perp}$ = 0, the proposed AH metal state and the nearly gapless semimetal state are competing orders with similar energy. Experimentally, due to limited sample quality, we cannot rule out the possibility that the observed AH metal behavior with finite longitudinal resistivity and non-quantized Hall resistivity arises from a nearly gapless IQAH state.

**Unconventional insulating-like transport at $\nu \approx$ -2.5**

From the $\nu$-$D$-$\rho_{xx}$ maps (Fig. 1c, Extended Data Fig. 2), we noticed that in addition to the insulating phases observed at integer fillings under relatively large $D$-fields, $\rho_{xx}$ is also significantly enhanced around $\nu \approx$ -2.2 to -2.7 near $D$ = 0 at low temperatures. Figure 4a presents the temperature dependence of $\rho_{xx}$ at $\nu \approx$ -2.5 under various $D$-fields. At large $|D|$ values, the temperature dependence of $\rho_{xx}$ resembles a metallic phase, and it follows $T^2$ dependence at low temperatures (Extended Data Fig. 11), consistent with the Fermi liquid behavior. With decreasing $|D|$, the temperature dependence of $\rho_{xx}$ gradually deviates from Fermi liquid behaviors. Remarkably, the temperature dependence of $\rho_{xx}$ becomes nonmonotonic when $|D|$ < 150 mV/nm, with $\rho_{xx}$ minimums appearing around $T \approx$ 5-10 K. Specifically, at high temperatures, $\rho_{xx}$ exhibits a metallic behavior; however, at low temperatures, $\rho_{xx}$ increases rapidly with decreasing $T$, roughly following a linear function as $1/T$ below 5 K, as shown in Fig. 4b.

The enhanced $\rho_{xx}$ at low temperatures around $\nu \approx$ -2.5 can be suppressed by applying a small $B_{\perp}$. Figure 4c illustrates the $\rho_{xx}$ versus $B_{\perp}$ at several $\nu$ between $\nu$ = -2 to -3 and $D \approx$ 0. It can be seen that the region with enhanced $\rho_{xx}$ at zero magnetic field shows strong negative out-of-plane magnetoresistance, up to 800% within $B_{\perp}$= 1 T at $\nu$ = -2.5.

Such magnetoresistance can be roughly fitted by $\rho_{\mathrm{xx}}(B_{\perp}) = a/B_{\perp} - b/B_{\perp}^2 + c$ (Extended Data Fig. 12). In contrast, $\rho_{\mathrm{xx}}$ at $\nu$ = -2 and -3 shows weak dependence on $B_{\perp}$. Figure 4d further demonstrates the temperature dependence of the out-of-plane magnetoresistance at $\nu \approx$ -2.5. The strong magnetoresistance is suppressed at elevated temperatures. At $B_{\perp}$= 1 T, we found the temperature dependence of $\rho_{\mathrm{xx}}$ at $\nu \approx$ -2.5 becomes metallic rather than exhibiting the nonmonotonic behavior mentioned above (Extended Data Fig. 12).

The $1/T$ temperature dependence in resistance are unlikely due to conventional Kondo scattering, which gives logarithm upturn resistivity at low temperature[61]. Regarding magnetoresistance behavior, Kondo scattering may however provide with negative magnetoresistance due to full polarization of electron's spin and consequent suppression of spin-flip channel of scattering. Nevertheless, studies[62-65] of Kondo effect suggest $-B^2$ scaling at small magnetic field and $B^{-4}$ scaling for large fields, contradicting the observed $\frac{a}{B} - \frac{b}{B^2}$ behavior. Therefore, the experimental observations may relate to scattering from correlated magnetic moments rather than from isolated magnetic moments. Indeed, previous studies[66] of frustrated itinerant magnets have pointed out that the non-saturating $1/T$ resistance at low temperature indicates liquid-like ground states and spin fluctuation-induced scattering, which get enhanced at low temperature but suppressed at high magnetic field (Methods). On the other hand, the insulating-like transport behavior at $\nu \approx$ -2.5 at $B$ = 0 may also originate from an incipient ferromagnetic insulating state. Further experimental and theoretical studies are needed to affirmatively determine the nature of this phase.

## Discussions and conclusions

In quantum Hall physics, the second Landau Level (LL) has attracted great interest. It not only hosts conventional odd-denominator fractional quantum Hall (FQH) states, but also the even-denominator FQH states[67] at LL filling factors 5/2 and 7/2. The former can be described by the Laughlin wavefunction or non-interacting composite Fermion model[68], and the latter is considered to be special FQH states with non-Abelian properties[69,70]. Moreover, the second LL exhibits subtle competitions between FQH states and other strong-correlation-induced nematic phases. For example, the 5/2 FQH state can be tuned into a stripe phase by applying a $B_{\parallel}$ [71,72] or hydrostatic pressure[73].

In a topological flat band, there are in principle much more quantum phases arising from the interplay between strong correlations and topology, which compete with the FQAH states[27-58]. Nevertheless, recent theoretical studies[49-56] have pointed out the possibility for realizing non-Abelian anyons in the half-filled second moiré flat band. Our experiment establishes the topological nature of the second moiré band in approximately 3° t$MoTe_2$, demonstrating a series of correlated and topological phases formed in this band. This opens the door to further explore more exotic quantum states, such as the non-Abelian FQAH state, in the second moiré flat band.

## Methods

### Device fabrications

We fabricated $tMoTe_2$ devices with a triple-gated geometry, as detailed in Ref. 10. In brief, the global $Si/SiO_2$ (285 nm) gate is used to achieve heavy hole doping in the contact regions, where metallic $TaSe_2$ directly touched with the $tMoTe_2$ layers. The top and bottom graphite gates are used to independently control the *n* and *D* in the $tMoTe_2$ channel. Atomically thin flakes of graphite, *h*-BN, $MoTe_2$, $TaSe_2$ were exfoliated from bulk crystals, and assembled into the desired stack using a standard dry transfer method[74] with polycarbonate stamps inside a nitrogen-filled glove box. The twisted angle between the two layers of $MoTe_2$ monolayer is controlled by a mechanic rotator. We used commercial 2H-$MoTe_2$ and 2H-$TaSe_2$ bulk crystals from HQ Graphene. The entire stack was released onto the $Si/SiO_2$ substrate with prepatterned Ti/Au (5 nm/20 nm) electrodes. The stack was further processed into a standard Hall bar geometry by standard *e*-beam lithography and reactive ion etching methods.

### Calibration of moiré filling factors

The device geometry allows us to independently tune the carrier density $\left(n = \frac{c_t V_t + c_b V_b}{e} + n_0\right)$ and the vertical electric displacement field $\left(D = \frac{c_t V_t - c_b V_b}{2\varepsilon_0} + D_0\right)$ in $tMoTe_2$ by applying top graphite gate voltage $V_t$ and bottom graphite gate voltage $V_b$. Here, $\varepsilon_0$, $c_t$, $c_b$, $n_0$ and $D_0$ denote the vacuum permittivity, geometric capacitance of the top graphite gate, geometric capacitance of the bottom graphite gate, intrinsic doping and the built-in electric field, respectively. The value of $c_t$ and $c_b$ are mainly determined by measuring quantum oscillations under large $B_\perp$ (Extended Data Fig. 1), and double checked by measuring the thickness of $h$-BN layers. We convert $n$ to moiré filling factor $v$ using the density difference between a series of correlated and/or topological quantum states with prominent $\rho_{xx}$ and $\rho_{xy}$ features.

## Transport measurements

Electrical transport measurements above 1.6 K were performed in a closed-cycle $^4$He cryostat (Oxford TeslatronPT) equipped with a 12 T superconducting magnet. Measurements below 1.6 K were performed in a top-loading dilution refrigerator (Oxford TLM, nominal base temperature about 15 mK) equipped with an 18 T superconducting magnet. The sample is immersed in the $^3$He-$^4$He mixtures during the measurements. A rotating probe was used in order to perform angle-dependent magnetoresistance measurements. No low-temperature filters were mounted on the rotating probe. The base electron temperature for Van der Waals devices using this particular probe is roughly estimated to be around 300-400 mK, based on prior experience with measuring superconductivity in graphene devices[75]. Electrical transport measurements were conducted using standard low-frequency lock-in techniques. The bias current is limited within 2 nA to avoid sample heating and disturbance of fragile quantum states. Voltage pre-amplifiers with 100 MΩ impedance were used to improve the signal-to-noise ratio. The contact resistance of device A and B is summarized in Extended Data Fig. 1. In general, the contact resistance is below 100 kΩ at low temperatures as long as the moiré filling factor $v$ is above -0.5. Below filling factor -0.5, the contact resistance increases rapidly as hole-doping density decreases.

Finite longitudinal-transverse coupling occurs in our devices that mixes the longitudinal resistance $R_{xx}$ and Hall resistance $R_{xy}$. To correct this effect, we used the standard procedure to symmetrize $\left(\frac{R_{xx}(B)+R_{xx}(-B)}{2}\right)$ and antisymmetrize $\left(\frac{R_{xy}(B)-R_{xy}(-B)}{2}\right)$ the measured $R_{xx}$ and $R_{xy}$ under positive and negative magnetic fields to obtain accurate values of $R_{xx}$ and $R_{xy}$, respectively. The longitudinal resistivity $\rho_{xx}$ is derived by $\rho_{xx} = R_{xx}\frac{W}{L}$, where $W$ is the Hall bar width and $L$ is the separation between voltage probes. The Hall resistivity $\rho_{xy}$ equals to $R_{xy}$ in two-dimensional case.

## MCD measurements

The MCD measurements were performed in reflection geometry in a home-built confocal optical microscope system based on a closed-cycle helium cryostat (base temperature 1.6 K) equipped with a superconducting magnet (9 T). A superluminescent light emitting diode with peak wavelength of 1070 nm and full-width at half-maximum bandwidth of 90 nm was used as the light source. The output of the diode was coupled to a single-mode fiber and focused onto the device under normal incidence by a low-temperature microscope objective [Attocube, 0.8 numerical aperture (NA)]. The spot size of the excitation light, defined as the FWHM of the diffraction-limited beam spot, can be calculated as $\mathrm{FWHM} = 0.51\frac{\lambda}{\mathrm{NA}}$ ~0.7 μm. A combination of a linear polarizer and an achromatic quarter-wave plate was used to generate the left ($\sigma^+$) and right ($\sigma^-$) circularly polarized light. The incident intensity on the sample was kept below $30\,\mathrm{nW}\,\mu\mathrm{m}^{-2}$ to avoid heating. The reflected light of a given helicity was spectrally resolved by a spectrometer coupled to a liquid nitrogen-cooled InGaAs one-dimensional array sensor (Princeton Instruments PyLoN-IR 1.7). The MCD spectrum is defined as $\frac{R^+-R^-}{R^++R^-}$, where $R^+$ and $R^-$ denote the reflection intensity of the left and right circularly polarized light. To analyze the MCD as a function of tuning parameters such as $\nu$, $D$, $B_\perp$ and $T$, we integrate the MCD modulus over a fixed spectral range, which covers most of the spectrum range of MCD signal. The integrated MCD (referred to simply as the MCD below) reflects the difference in occupancy between the K and K′ valleys in $MoTe_2$. Because of spin-valley locking, the signal is proportional to the out-of-plane magnetization[76-79]. In the MCD map measurements, the results are averaged by positive and negative $B_\perp$, i.e., $MCD = \frac{MCD(+B_\perp)-MCD(-B_\perp)}{2}$.

## Continuum model and Hartree-Fock calculation

We employ the continuum model, which includes the first-harmonic, second-harmonic and displacement field terms to describe moiré band topology. Given that the K and K' valleys are connected through time-reversal symmetry, analyzing one valley is sufficient to infer the entire band topology. Constrained by the three-fold rotational symmetry ($C_{3z}$), we derive the two-band $\boldsymbol{k}\cdot\boldsymbol{p}$ Hamiltonian as the following form:

$$H = \begin{bmatrix} -\frac{(k-K_t+eA)^2}{2m^*} + \Delta_t(r) - \frac{\epsilon}{2} & \Delta_T(r) \\ \Delta_T^\dagger(r) & -\frac{(k-K_b-eA)^2}{2m^*} + \Delta_b(r) + \frac{\epsilon}{2} \end{bmatrix}$$

The intralayer and interlayer potential terms, strain induced gauge field take the forms:

$$\Delta_t(\boldsymbol{r}) \quad = 2V_1 \sum_{i=1,3,5} \cos(\boldsymbol{g}_i^1 \cdot \boldsymbol{r} + l\phi_1) + 2V_2 \sum_{i=1,3,5} \cos(\boldsymbol{g}_i^2 \cdot \boldsymbol{r})$$
$$\Delta_T \quad = w_1 \sum_{i=1,2,3} e^{-i\boldsymbol{q}_i^1 \cdot \boldsymbol{r}} + w_2 \sum_{i=1,2,3} e^{-i\boldsymbol{q}_i^2 \cdot \boldsymbol{r}}$$
$$A(\boldsymbol{r}) \quad = A(\boldsymbol{a_2} \sin(\boldsymbol{G_1} \cdot \boldsymbol{r}) - \boldsymbol{a_1} \sin(\boldsymbol{G_3} \cdot \boldsymbol{r}) - \boldsymbol{a_3} \sin(\boldsymbol{G_5} \cdot \boldsymbol{r}))$$

Here, *k* is the momentum measured from the Γ point of a single-layer $MoTe_2$, $K_t$ and $K_b$ represent the high symmetry momentum of the top and bottom layers, respectively. The $\boldsymbol{G_{1,3,5}}$ are the moiré reciprocal vectors along three different in-plane directions, and the $\boldsymbol{a_{1,2,3}}$ are the corresponding moiré lattice vectors in real space. The terms $\boldsymbol{g_i^1}$ and $\boldsymbol{g_i^2}$ represent the momentum differences between the nearest and second-nearest plane wave bases within the same layer, respectively. Similarly, $\boldsymbol{q}_i^1$ and $\boldsymbol{q}_i^2$ represent the momentum differences between the nearest and second-nearest plane wave bases across different layers. $\Delta_t(r)/\Delta_b(r)$ denotes the layer-dependent moiré potential, and $\Delta_T(r)$ is the interlayer tunneling term. $A$ is the strain-induced gauge field and $\epsilon$ is the displacement field term, which break the layer-exchange symmetry ($C_{2y}T$). In total, there are six independent parameters. Then, we fit the density functional theory band structures under dDsC van der Waals correction, and get the following parameter[34]: $m^*$ = 0.62 $m_e$, $V_1$ = 10.3 meV, $V_2$ = 2.9 meV, $w_1$ = -7.8 meV, $w_2$ = 6.9 meV, $\varphi_1$ = -75°, $\Phi/\Phi_0$ = 0.737. The continuum model band structure is shown in Extended Data Fig. 8a. Employing these parameters, we diagonalize the Hamiltonian and calculate the density of states (DOS) according to:

$$D(E) = \frac{N_e}{(2\pi)^2} \sum_n \int_{\mathrm{BZ}} \delta(E - \epsilon_{n,\mathbf{k}}) d^2k$$

where $N_e$ represents the band occupancy and $\delta(E - \epsilon_{n,\mathbf{k}})$ is the delta function centered at the energy *E*. The result is depicted in Extended Data Fig. 8b. To approximate the delta function, we use a Gaussian broadening:

$$\delta(E - E_n) \approx \frac{1}{\sqrt{2\pi\sigma^2}} \exp\left(-\frac{(E - E_n)^2}{2\sigma^2}\right)$$

Here, $\sigma$ is the smearing parameter, and we set to 0.5 meV during the calculation. By analyzing the peaks of the DOS between the minimum and maximum energy interval of the second band, we pinpoint the energy position of the Van Hove singularity and draw the Fermi surface at that position. In Extended Data Fig. 8d we show three ordinary van Hove singularities (V) and one high-order van Hove singularity (H) at filling factor $\nu$ = -3.

The Hartree-Fock Hamiltonian is given by:

$$H = \sum_{s,s'} \left[ -\sum_{Q\in Q_0} \frac{(k-Q)^2}{2m^*} + \sum_{Q,Q'\in Q_0; j=1}^{3} w_1 \delta_{Q',Q\pm q_i^1} + \sum_{Q,Q'\in Q_0; j=1}^{3} w_2 \delta_{Q',Q\pm q_i^2} \right.$$

$$\left. + \sum_{l=\pm} \sum_{Q,Q'\in Q_l; i=1,3,5} V_1 \left( e^{il\phi} \delta_{Q',Q+g_i^1} + e^{-il\phi} \delta_{Q',Q-g_i^1} \right) + \sum_{Q,Q'\in Q_0} V_2 \delta_{Q,Q'\pm g_i^2} \right] \delta_{ss'}$$

$$+ \sum_{s,s';Q,Q'\in Q_0} \frac{1}{A} \sum_{k',g'';Q''\in Q_0} \left[ \delta_{s,s'} \sum_{s''} V_{s,s''}(g'') \rho_{s'',Q''-g'';s'',Q''}(k') \right.$$

$$\left. -V_{s,s'}(k'-k-Q''+Q) \rho_{s',Q''-g'';s,Q''}(k') \right] \delta_{Q',Q-g''}$$

Throughout the calculations, we take $\theta = 3.15°$ We use $Q_\pm = \pm q_1^1 + G_m$, where $q_1^1$ measures the momentum difference between $k_t$ and $k_b$, and $G_m = mg_1^1 + ng_3^1$ represents the moiré reciprocal lattice. We define $Q_0 = Q_+ \oplus Q_-$. Elements in $Q_0$ label a set of plane wave basis, where $Q_+$ corresponds to the top layer and $Q_-$ to the bottom layer. We define $\rho_{s'Q'sQ}(k) = \langle \psi_{HF} | c^\dagger_{s,k-Q} c_{s',k-Q'} | \psi_{HF} \rangle$. We use a screened Coulomb potential: $V_{s,s'}(q) = \frac{e^2}{2\epsilon\epsilon_0} \frac{1}{\sqrt{q^2+(\frac{1}{r_0})^2}}$. The area, $A$ is contingent on the k-mesh size in calculations, typically a $15 \times 15$ Monkhorst-Pack Brillouin zone mesh, and the lowest 20 shells for the plane wave basis are considered.

In Extended Data Fig. 9a, we present the phase diagram of the direct band gap as a function of the gating potential, with time-reversal symmetry enforced. The filling factors for the two valleys are set to $\nu_\uparrow = -1$ and $\nu_\downarrow = -1$. The influence of the gating potential on the Hamiltonian is expressed as $\sum_{l=\pm} \sum_{Q\in Q_l} \frac{l\Delta}{2} \delta_{Q',Q} \delta_{s,s'}$, where $\Delta$ represents the gating potential. Unless otherwise stated, the dielectric constant is fixed at $\epsilon = 30$ throughout the following calculations. As $\Delta$ increases from 0, we observe the gap close and reopen, signaling a phase transition from an IQSH phase to a trivial insulator. With no gating field applied, the band exhibits IQSH features, with the top Chern band of $K$ valley showing a Chern number of 1, and the top Chern band of $K'$ valley displaying a Chern number of -1. The gap closure at $K'$ occurs at $\Delta_c = 22.5\text{meV}$. After the gap closure, the top two bands become trivial, aligning with experimental observations that a gating field can drive the IQSH state to a trivial state.

In Extended Data Fig. 9b and 9c, we show the band structure at a total filling factor $\nu = -3$, under varying interaction strength. Without applying magnetic field, two potential candidate states are identified: an anomalous Hall (AH) metal state at $\epsilon = 20$, converging from an unpolarized initial state, and a nearly gapless semimetal state with a quantized Hall response (equivalent to a nearly gapless IQAH state) at $\epsilon = 30$. The band structure of the AH metal displays features consistent with experimental results. The Chern number sequence for the $K$ valley bands, from top to bottom, is (1, -2), while for the $K'$ valley, the sequence is (0, -2), as illustrated in the schematic band structure in Extended Data Fig. 9b. The -2 Chern number for the second band in the top in $K$ valley originates from band touching of the second and third band when $\epsilon$ slightly exceeds 20, transforming a -3 Chern number through band touching along three $\Gamma - K$ lines, changing it to -2 from 1 calculated in the non-interacting limit. Here, two completely filled bands contribute a total Chern number of 1, while two partially filled bands, with both Chern number -2, offset the positive contribution and result in a negative contribution in total. This is consistent with the experimental observation that the sign of the anomalous Hall signals around $\nu = -3$ is opposite to that of the $\nu = -1$ IQAH state.

On the other hand, the nearly gapless state at $\epsilon = 30$ is valley polarized, having filling factors for each valley $\nu_{\uparrow} = -2$ and $\nu_{\downarrow} = -1$, as shown schematically in Extended Data Fig. 9c. The Chern numbers of bands, from top to bottom at $K$, are (1, -1) and -1 at $K'$. This state has a small direct band gap at $\Gamma$, approaximately 0.57 meV. Given the relatively small gap and limited sample quality, we cannot rule out the possibility of the observed AH metal behavior at $\nu = -3$ is actually arising from this state.

The nearly gapless semimetal state can be adiabatically connected to a Chern insulator state with larger gap by adding an effective Zeeman field. When an external magnetic field is applied, it is described by the term $\sum_{s=\pm} \sum_{Q \in Q_0} \frac{1}{2} E_z s \delta_{Q',Q} \delta_{s,s'}$, where $E_z$ represents the strength of the Zeeman field. As shown in in Extended Data Fig. 9d, the magnetic field drives the system to a gapped state by shifting the energy between two valleys. This results in an increase in the gap to approximately 2.57 meV. The total Chern number of the state is -1, which aligns with the experimentally observed state with the same Chern number using Streda formula and $\rho_{xy}$ measurements.

### Spin fluctuation induced 1/*T* resistance

Low temperature scaling of resistivity and the magnetic field behavior (discussed below) could be the result of scattering at spin fluctuations. In this regard, resistivity behavior is controlled by spin structure factor $\langle S_k S_{-k} \rangle$:

$$\rho(T) \propto \int dk \langle S_k S_{-k} \rangle (1 - \cos \theta_{k,k+p})$$

Spin structure factor could be described with the help of short-ranged Ruderman-Kittel-Kasuya-Yosida (RKKY) type of Hamiltonian:

$$H_{RKKY} = -J \sum_{i,j} S_i \cdot S_j$$

so that thermal average for spin structure factor reads (see also Ref. 66),

$$\langle S_k S_{-k} \rangle = \frac{\int DS S_k S_{-k} e^{-\beta H_{RKKY}}}{\int DS e^{-\beta H_{RKKY}}} \propto 1 + \frac{cJ}{T},$$

where $c$ is a numerical constant. It should also be noted that the similar $1/T$ scaling was recently observed in $Yb_2Ti_2O_7$ quantum magnet[80], demonstrating strong spin fluctuations in low temperature domain $T \leq 10\ K$.

Regarding $B_\perp$ dependence in the domain of large magnetic fields, Zeeman single spin energy becomes dominant and destroys correlations between neighboring magnetic moments. So that, spin-spin correlation function becomes local in real space and is approximated by Brillouin function:

$$\langle S_k S_{-k} \rangle \approx \frac{1}{\tilde{B}} - \frac{1}{\tilde{B}^2}, \qquad \tilde{B} = \frac{\mu_B B}{k_B T}.$$

**Acknowledgement**

We thank Fengcheng Wu, Zhao Liu, Xiaoyan Xu and Mingpu Qin for helpful discussions. This work is supported by the National Key R&D Program of China (Nos. 2022YFA1402702, 2022YFA1405400, 2021YFA1401400, 2021YFA1400100, 2022YFA1402404, 2019YFA0308600, 2022YFA1402401, 2020YFA0309000), the National Natural Science Foundation of China (Nos. 12350403, 12174249, 12174250, 12141404, 92265102, 12374045), the Innovation Program for Quantum Science and Technology (Nos. 2021ZD0302600 and 2021ZD0302500), the Natural Science Foundation of Shanghai (No. 22ZR1430900). T.L., S.J. and X.L. acknowledge the Shanghai Jiao Tong University 2030 Initiative Program. X.L. acknowledges the Pujiang Talent Program 22PJ1406700. T.L. and S.J. acknowledges the Yangyang Development Fund. Yang Zhang acknowledges support from AI-Tennessee and Max Planck partner lab grant. K.W. and T.T. acknowledge support from the JSPS KAKENHI (Nos. 21H05233 and 23H02052) and World Premier International Research Center Initiative (WPI), MEXT, Japan. A portion of this work was carried out at the Synergetic Extreme Condition User Facility (SECUF).

**Author contributions**

T.L., S.J. and X.L. designed and supervised the experiment. J.X, F.L, F.X, X.C and Z.S. fabricated the devices. F.X., Z.S. and J.L. performed the transport measurements. X.C and J.X. performed the optical measurements. F.X., X.C., T.L., S.J. and X.L. analyzed the data. Yixin Zhang, N.M., N.P. and Yang Zhang performed theoretical studies. K.W. and T.T. grew the bulk hBN crystals. T.L. and Yang Zhang wrote the manuscript. All authors discussed the results and commented on the manuscript.

**Competing interests**

The authors declare no competing financial interests.

**Main Figures**

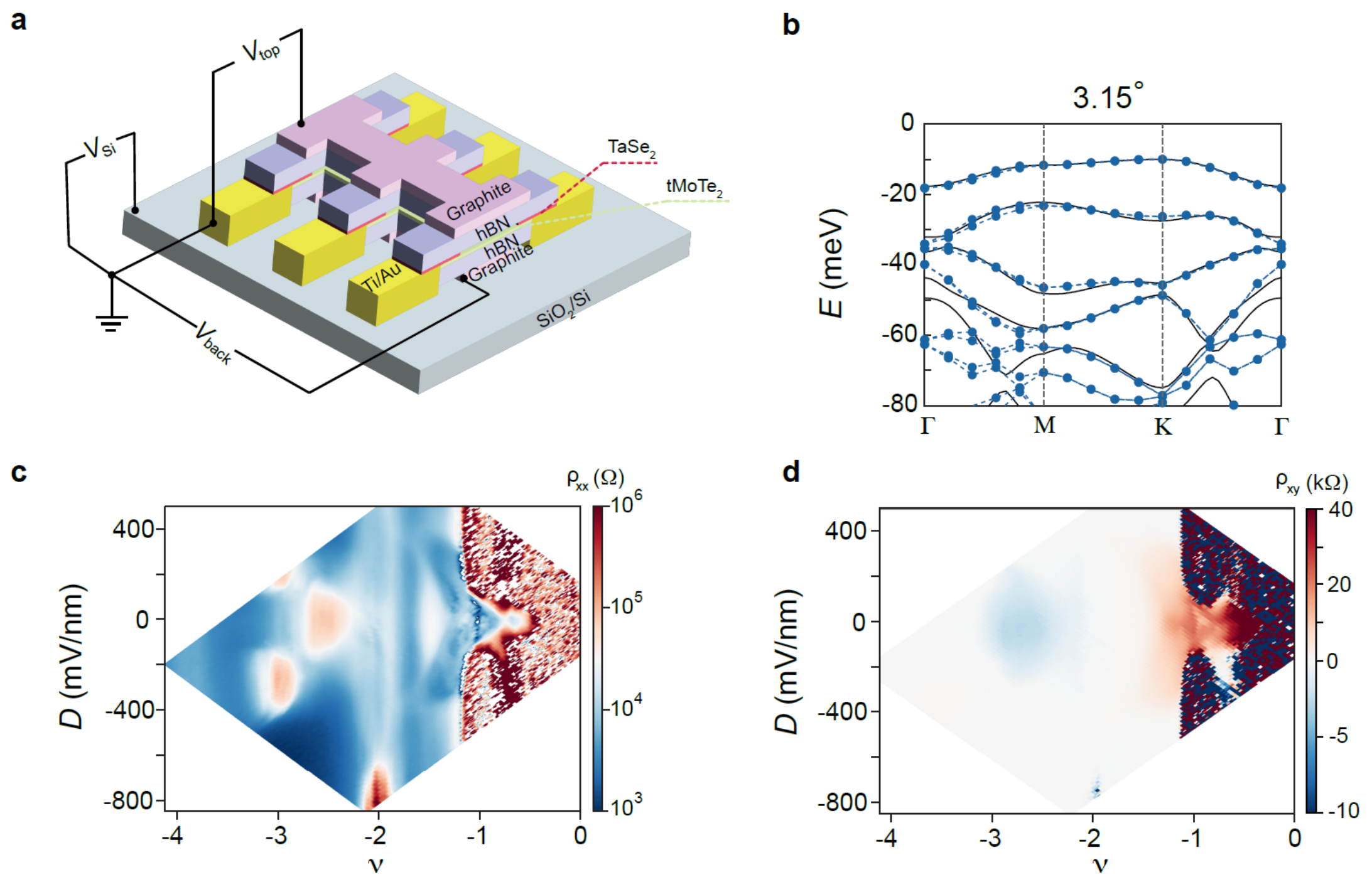

**Fig. 1. Characterization of 3.15° tMoTe2 devices. a**, Schematic of the triple-gated tMoTe2 device. **b**, Calculated band structure of 3.15° tMoTe2, where blue dots and black lines are DFT bands and continuum model bands, respectively. The first two moiré bands are well isolated at $D \approx 0$, and with Chern number +1. **c**,**d**, Longitudinal resistivity $\rho_{xx}$ (**c**) and Hall resistivity $\rho_{xy}$ (**d**) as a function of $\nu$ and $D$ for device A (3.15°). $\rho_{xx}$ is measured under zero magnetic field at 300 mK; $\rho_{xy}$ is the antisymmetrized results under an out-of-plane magnetic field $B_{\perp} = \pm$ 0.1 T at 1.6 K.

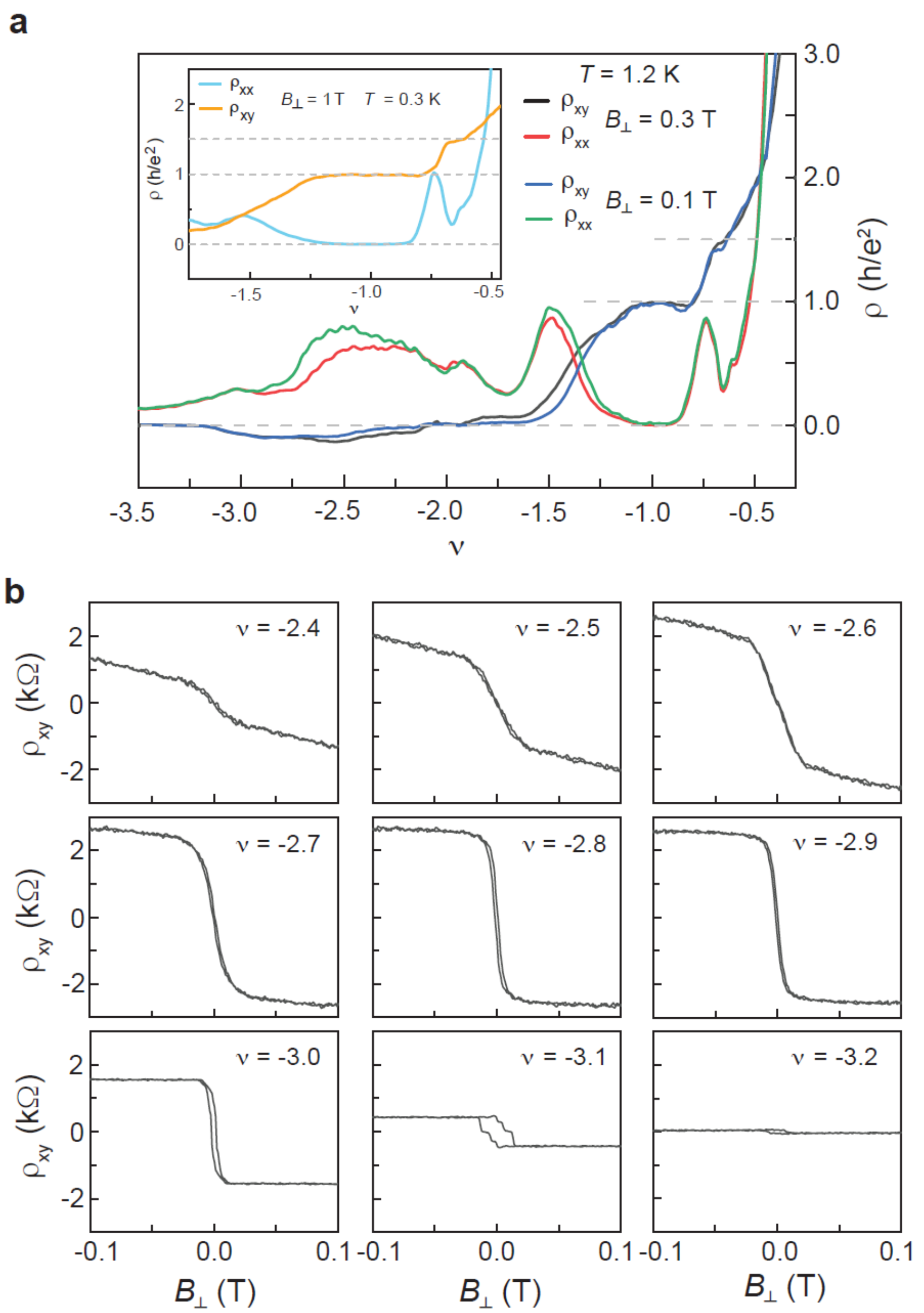

**Fig. 2. IQAH and FQAH effects in the first moiré band, and emergent magnetism in the second moiré band of device A (3.15°). a**, Symmetrized $\rho_{xx}$ and antisymmetrized $\rho_{xy}$ under $B_{\perp} = \pm\,0.1$ T and $\pm\,0.3$ T at $D \approx 0$ V/nm and $T = 1.2$ K. We observed an $\rho_{xy}$ plateau quantized to $h/e^2 \pm 1\%$ for the IQAH state at $\nu = -1$. As for the FQAH state at $\nu = -2/3$, we observed a weaker $\rho_{xy}$ plateau quantized to $3h/2e^2 \pm 2.5\%$ under $B_{\perp} = \pm 0.3$ T and $3h/2e^2 \pm 6\%$ under $B_{\perp} = \pm 0.1$ T. The inset shows similar measurements at $B_{\perp} = \pm 1$ T and $T = 0.3$ K. **b**, Antisymmetrized $\rho_{xy}$ versus $B_{\perp}$ at varying $\nu$ (from -2.4 to -3.2), measured at about 1.5 K.

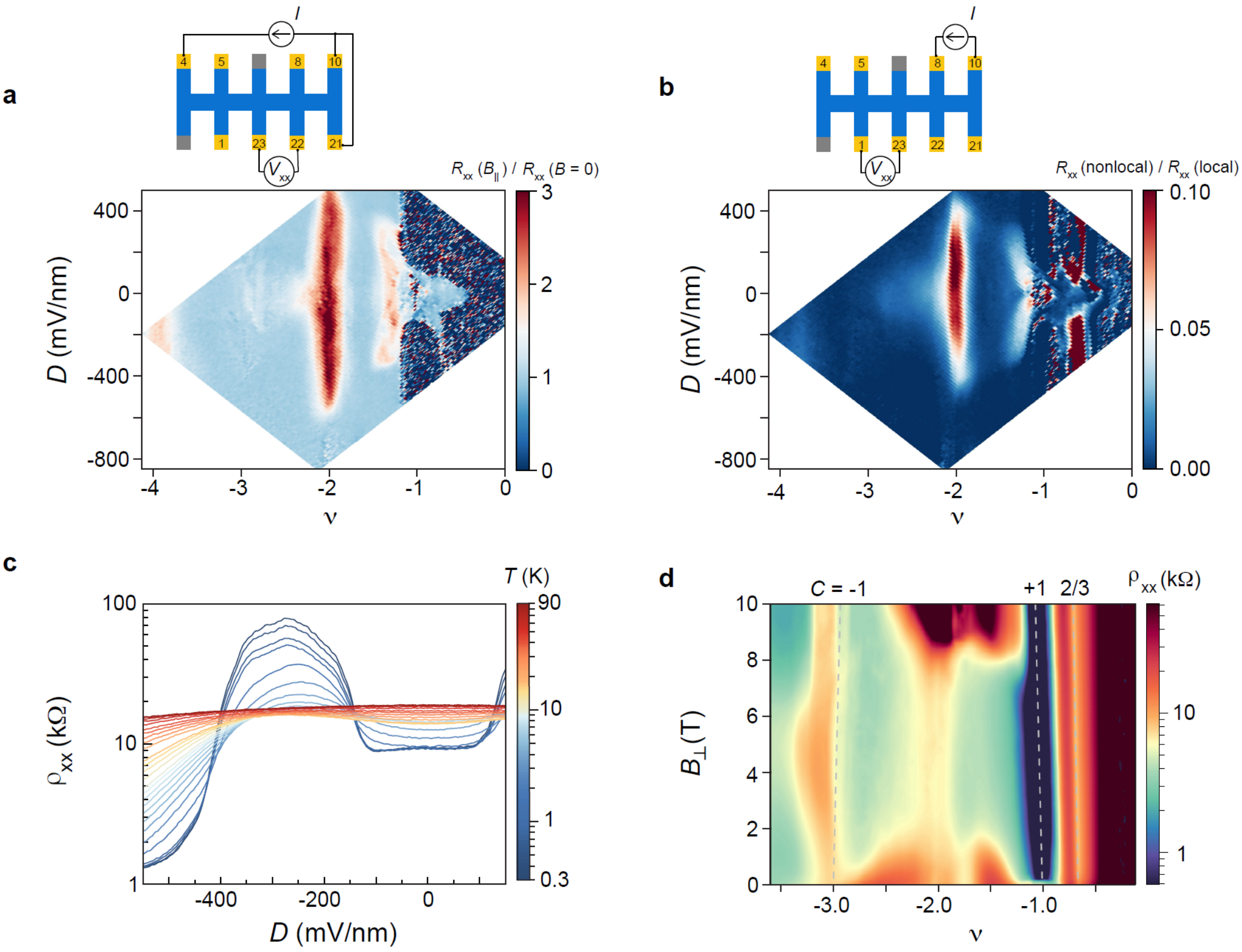

**Fig. 3. Topological quantum states formed at *ν* = -2, -3, and -4 in device A (3.15°).** **a**, The in-plane magnetoresistance $R_{xx}(B_{\|} = 0.3\ \mathrm{T})/R_{xx}(B = 0)$ versus *ν* and *D* measured at *T* = 300 mK. **b**, The ratio between nonlocal resistance and local resistance, versus *ν* and *D* measured at *T* = 1.6 K and $B_{\perp} = \pm\ 0.1$ T. The inset in **a** and **b** illustrates the measurement configurations, respectively. We noticed that both the in-plane magnetoresistance and the nonlocal signals can also be observed from *ν* ≈ -1.15 to -1.5 under certain *D* values, which requires further experimental and theoretical investigations. **c**, Temperature dependence of $\rho_{xx}$ versus *D* at *ν* = -3 from *T* = 300 mK to 90 K. **d**, $\rho_{xx}$ versus *ν* and $B_{\perp}$ at *T* = 1.6 K and *D* ≈ 0 V/nm. Dashed lines represent the expected dispersions based on Streda formula for the IQAH state at *ν* = -1 with *C* = 1, the FQAH state at *ν* = -2/3 with *C* = 2/3, and the Chern state emerged under $B_{\perp}$ at *ν* = -3 with *C* = -1, respectively.

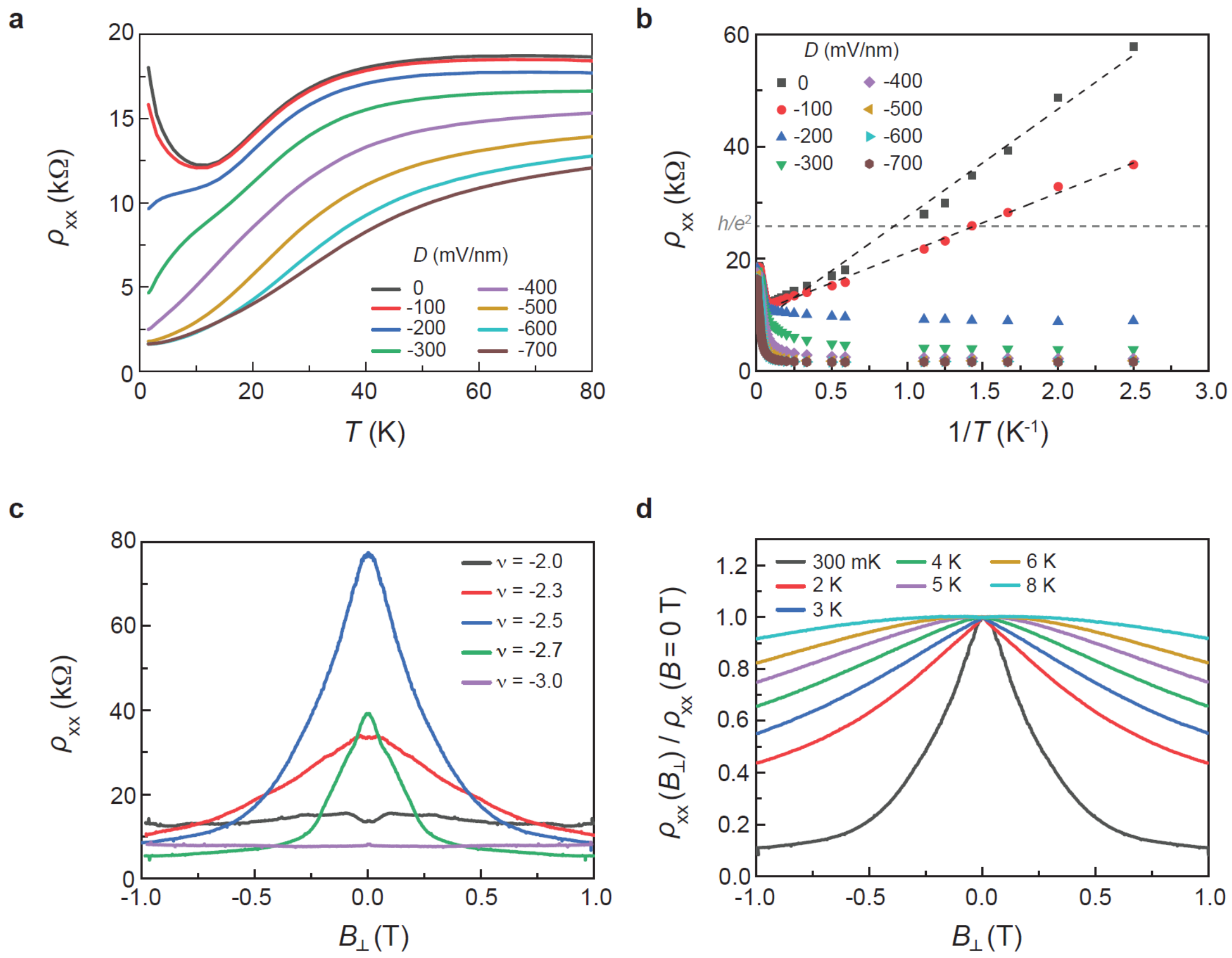

**Fig. 4. Transport behaviors at $\nu \approx$ -5/2 of device A (3.15°). a**, Temperature dependence ($T$ from 1.6 K to 80 K) of $\rho_{xx}$ at varying $D$ at $\nu \approx -5/2$. **b**, $\rho_{xx}$ versus $1/T$ at varying $D$, combining the data points measured at $^4$He cryostat ($T$ = 1.6 K to 80 K) and dilution refrigerator ($T$ = 300 mK to 900 mK). **c**, $\rho_{xx}$ versus $B_\perp$ at varying $\nu$ in the second moiré band, at $D \approx 0$ and $T$ = 300 mK. **d**, The out-of-plane magnetoresistance $\rho_{xx}(B_\perp)/\rho_{xx}(B = 0\ \mathrm{T})$ versus $B_\perp$ at $\nu \approx -5/2$ with varying temperature.

## Extended Data Figures

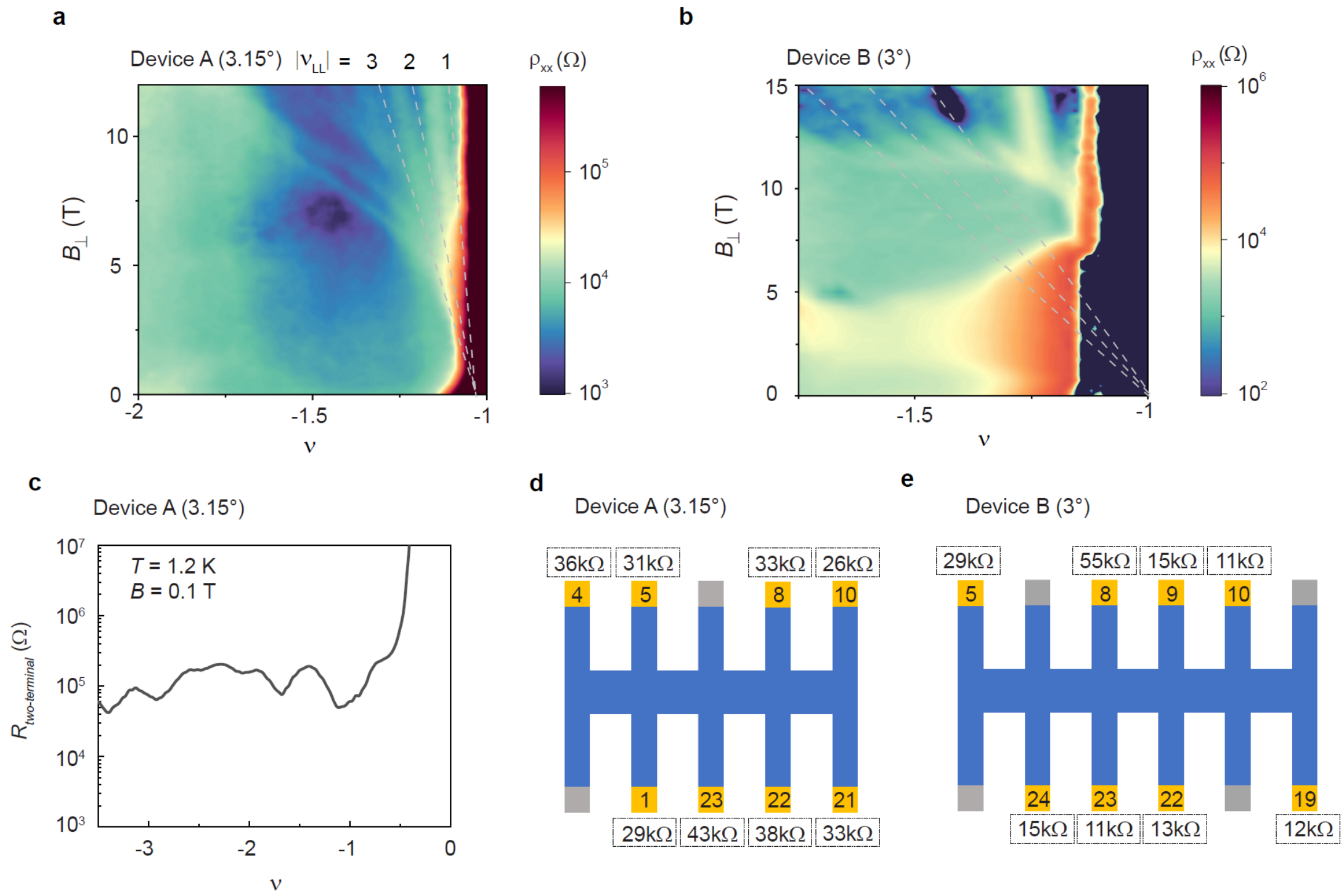

**Extended Data Fig. 1. Twist angle calibrations and contact resistance. a**, $\rho_{xx}$ versus $\nu$ and $B_\perp$ at $T$ = 300 mK and $D \approx$ -275 mV/nm of device A. Clear quantum oscillations can be observed around $\nu$ = -1 when $B_\perp$ above about 6 T, corresponding to a quantum mobility about 1500 cm$^2$/V·s. The Landau Level filling factor $\nu_{LL}$ can be determined from the Hall resistance, which allows us to derive the gate geometric capacitance accurately. **b**, $\rho_{xx}$ versus $\nu$ and $B_\perp$ at $T$ = 300 mK and $D \approx$ -410 mV/nm of device B. The twisted angles for device A and B are determined to be 3.15° ± 0.1° and 3.0° ± 0.1°, respectively. **c**, Two-terminal resistance as a function of $\nu$ at $T$ = 1.2 K and $B_\perp$= 0.1 T around $D$ = 0. It is measured with contact 21 and contact 4 of device A with a constant bias current of approximately 1 nA. The measured two-terminal resistance includes contributions from both the sample resistance and the contact resistance of the two contacts used as the source and drain for transport measurements. Given that the contact resistance is in general high in TMDc devices, we can reasonably assume that it dominates the two-terminal resistance, except when the sample is tuned into an insulating state. **d**,**e**, Contact resistance in device A (**d**) and device B (**e**) at 1.7 K. For device A, the contact resistance is measured at $\nu$ = -1.4 and $D$ = -116 mV/nm. For device B, the contact resistance is measured at $\nu$ = -4.2 and $D$ = -30 mV/nm. During the measurement of contact resistance, only the specific contact is biased with a 1 mV dc voltage while all other contacts are grounded.

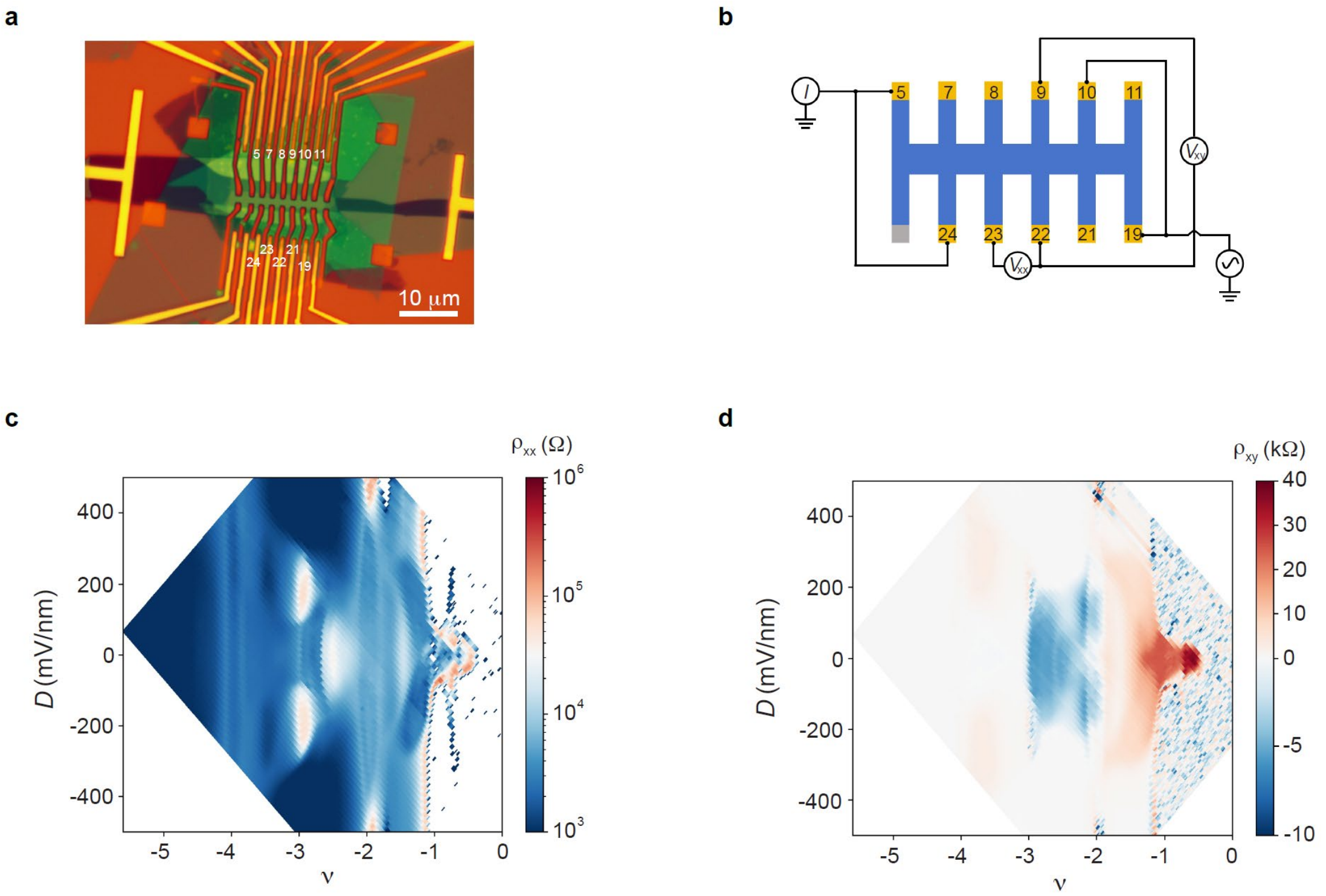

**Extended Data Fig. 2. Characterizations of device B (3.0°). a**, Optical image of device B (3.0°). **b**, The schematic transport measurement configuration. Contacts in **a** and **b** are labeled by numbers. **c**,**d**, Longitudinal resistivity $\rho_{xx}$ (**c**) and Hall resistivity $\rho_{xy}$ (**d**) as a function of $\nu$ and $D$. $\rho_{xx}$ is measured under zero magnetic field at 500 mK; $\rho_{xy}$ is the antisymmetrized results under an out-of-plane magnetic field $B_{\perp} = \pm$ 0.3 T at 500 mK. A built-in electric field $D_0$ about 50 mV/nm has been subtracted for device B (3.0°).

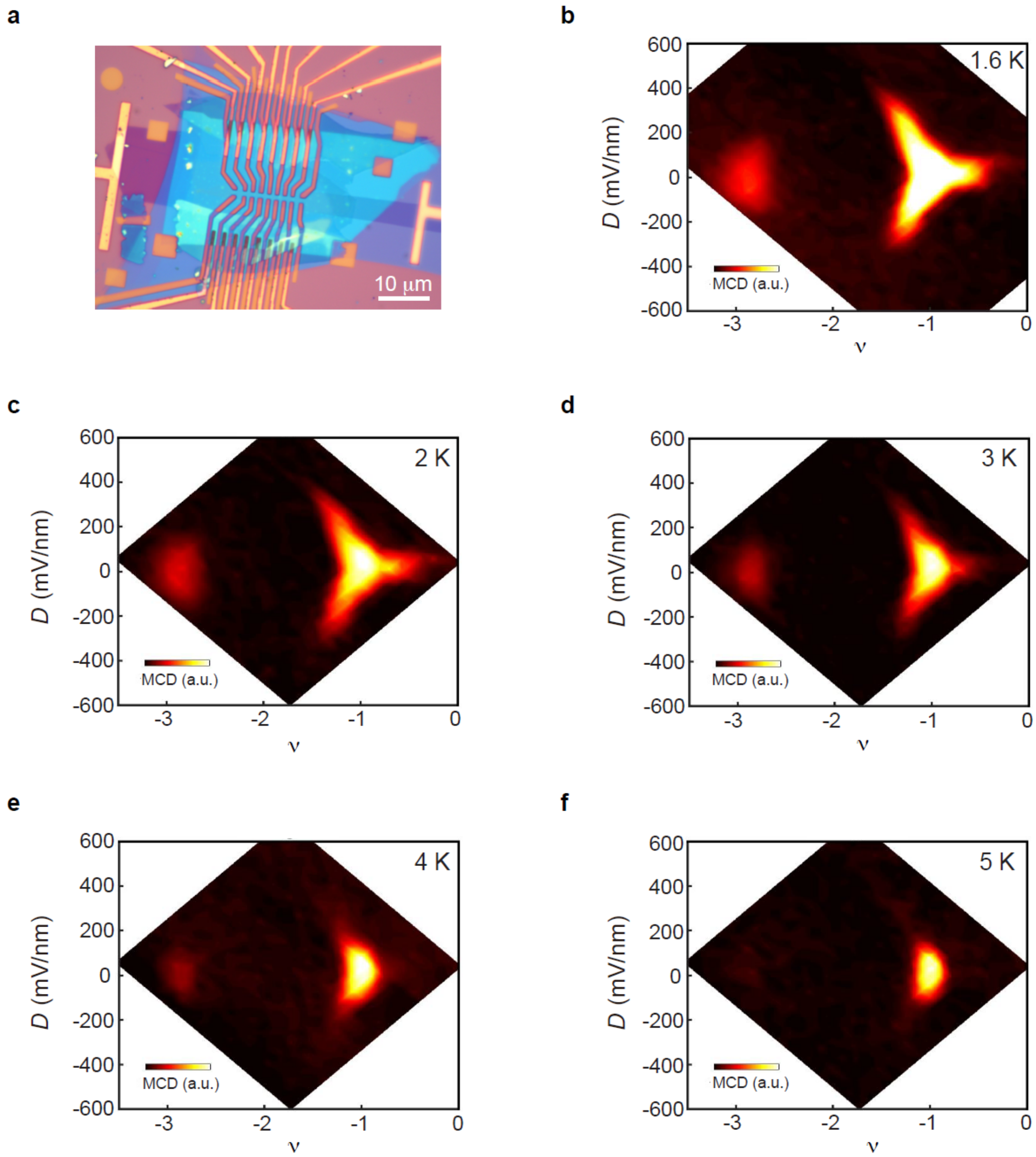

**Extended Data Fig. 3. Characterizations of device C (3.7°). a**, Optical image of device C (3.7°). **b-f**, MCD map as a function of $\nu$ and $D$ measured at $B_{\perp} \approx 10$ mT and $T$ = 1.6 K (**b**), 2 K (**c**), 3 K (**d**), 4 K (**e**), 5 K (**f**). Ferromagnetism in the second moiré band spans approximately from $\nu \approx$ -2.65 to -3.2. The maximum Cuire temperature of the second moiré band magnetism is about 5 K.

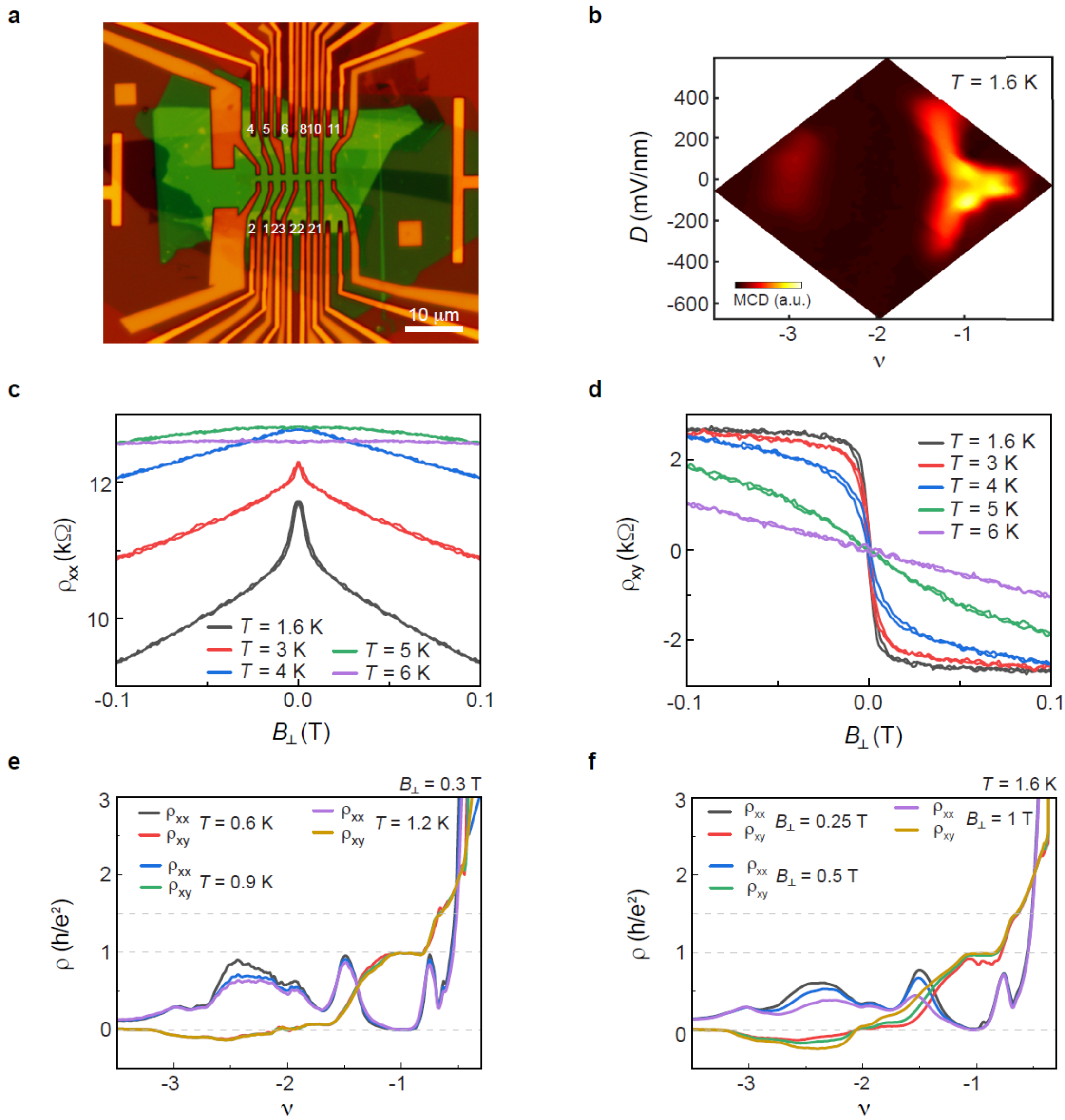

**Extended Data Fig. 4. More characterizations of device A (3.15°). a**, Optical image of device A (3.15°). Contacts are labeled by numbers. **b**, MCD map as a function of $\nu$ and $D$ measured at 1.6 K. **c**,**d**, Symmetrized $\rho_{xx}$ (**c**) and antisymmetrized $\rho_{xy}$ (**d**) at varying temperatures at $\nu \approx$ -2.8. **e**, Symmetrized $\rho_{xx}$ and antisymmetrized $\rho_{xy}$ under $B_{\perp} = \pm$ 0.3 T at $D \approx 0$ V/nm and $T = 0.6$ K, 0.9 K, 1.2 K, respectively. **f**, Symmetrized $\rho_{xx}$ and antisymmetrized $\rho_{xy}$ under at $D \approx 0$ V/nm and $T$ =1.6 K with $B_{\perp}$= 0.25 T, 0.5 T, 1 T, respectively.

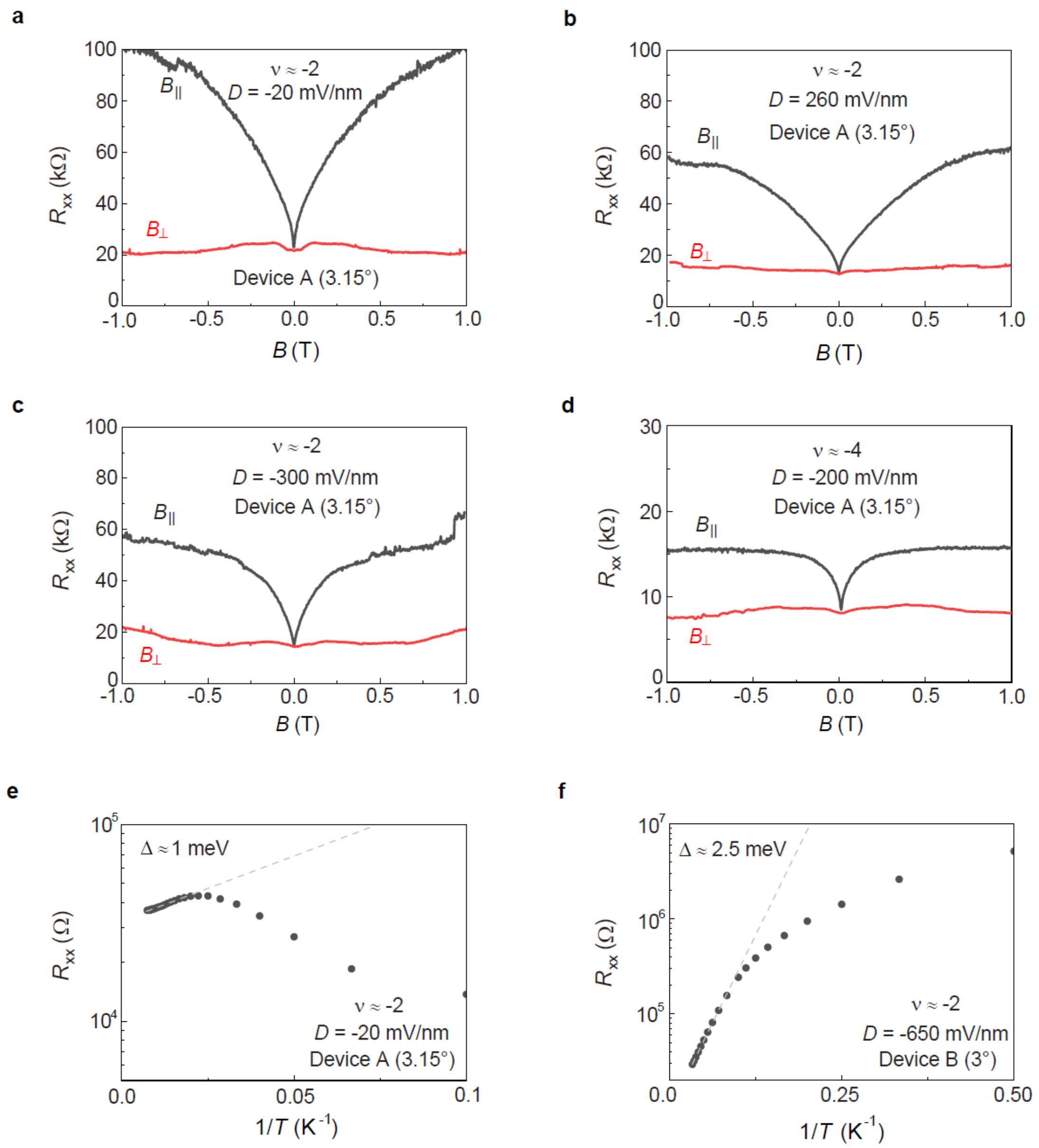

**Extended Data Fig. 5. Magnetoresistance at *ν* = -2 and -4 of device A (3.15°), and the transport gap fitting at *ν* = -2. a-c**, $R_{xx}$ versus $B_{\|}$ and $B_{\perp}$ at $\nu$ = -2 and $D$ = -20 mV/nm (**a**), $D$ = 260 mV/nm (**b**) and $D$ = -300 mV/nm (**c**). **d**, $R_{xx}$ versus $B_{\|}$ and $B_{\perp}$ at $\nu$ = -4 at $D$ = -200 mV/nm. The measurement temperature for **a-d** is 300 mK. **e**, Temperature dependence of $R_{xx}$ at $\nu$ = -2 and $D$ = -20 mV/nm of device A (3.15°). **f**, Temperature dependence of $R_{xx}$ at $\nu$ = -2 and $D$ = -650 mV/nm of device B (3.0°). Dashed lines in **e** and **f** are the fit to $\rho_{xx} \propto e^{-\Delta/2k_BT}$, with $\Delta$ and $k_B$ denoting the transport gap and Boltzmann constant, respectively. We observe that below about 40 K, the temperature dependence of $R_{xx}$ shown in **e** becomes metallic-like, presumably due to the interplay between bulk transport and edge transport of an IQSH insulator. Consequently, the thermal activation fitting range is very narrow in **e**, leading to a substantial uncertainty of the estimation for the real charge gap.

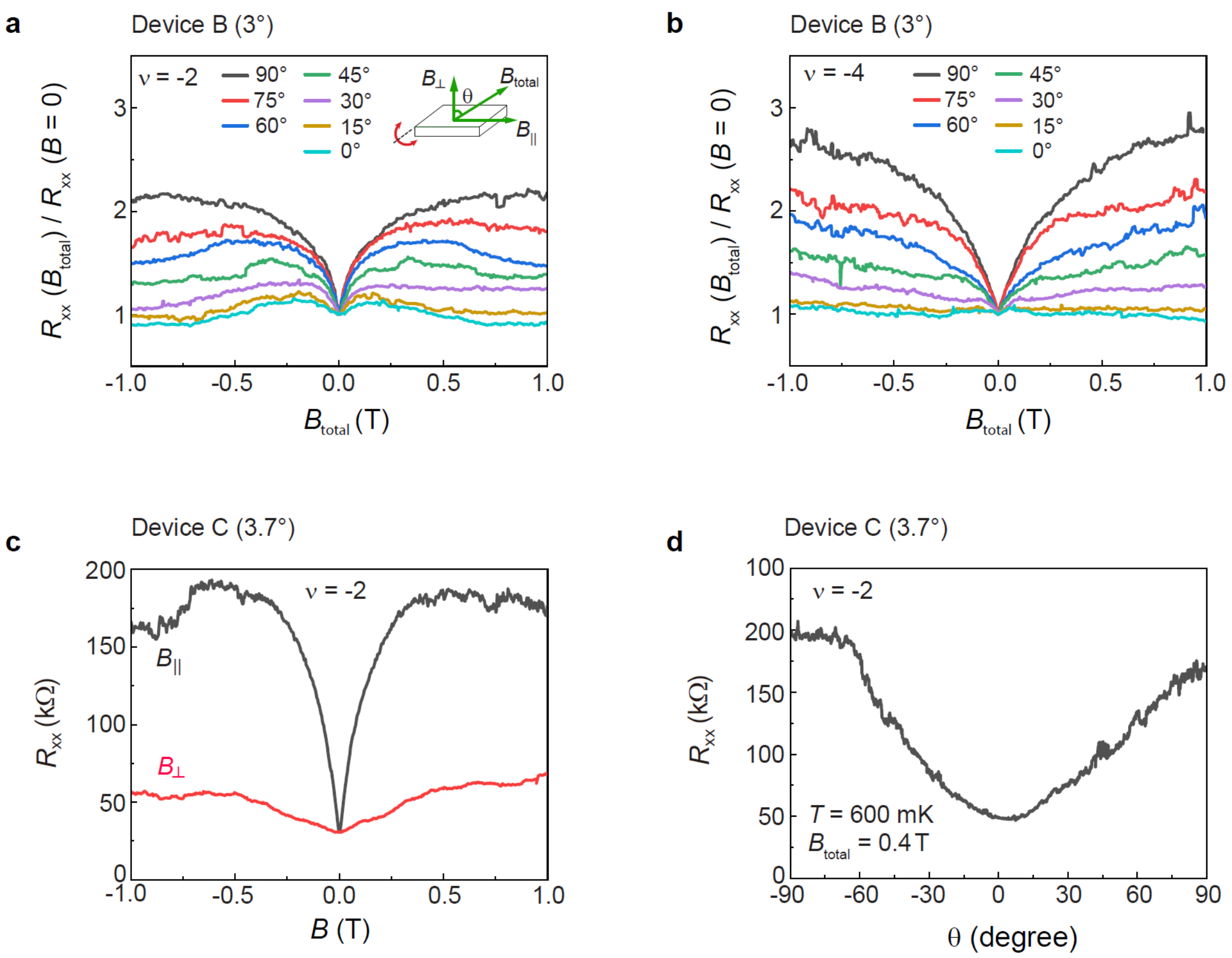

**Extended Data Fig. 6. Magnetoresistance at $\nu$ = -2 and -4 of device B (3.0°) and at $\nu$ = -2 of device C (3.7°). a**,**b**, Angle-dependent magnetoresistance at $\nu$ = -2 (**a**) and $\nu$ = -4 (**b**) of device B (3.0°) at 300 mK. The definition of the tilt angle $\theta$ is schematically illustrated in the inset of **a**, where $\theta$ = 90° corresponds to the in-plane magnetic field configuration. **c**, $R_{xx}$ versus $B_{\parallel}$ and $B_{\perp}$ at $\nu$ = -2 and $D \approx 0$ V/nm of device C (3.7°). **d**, $R_{xx}$ as a function of $\theta$ at $\nu$ = -2 and $D \approx 0$ V/nm of device C. The applied magnetic field $B_{total}$ is fixed at 0.4 T.

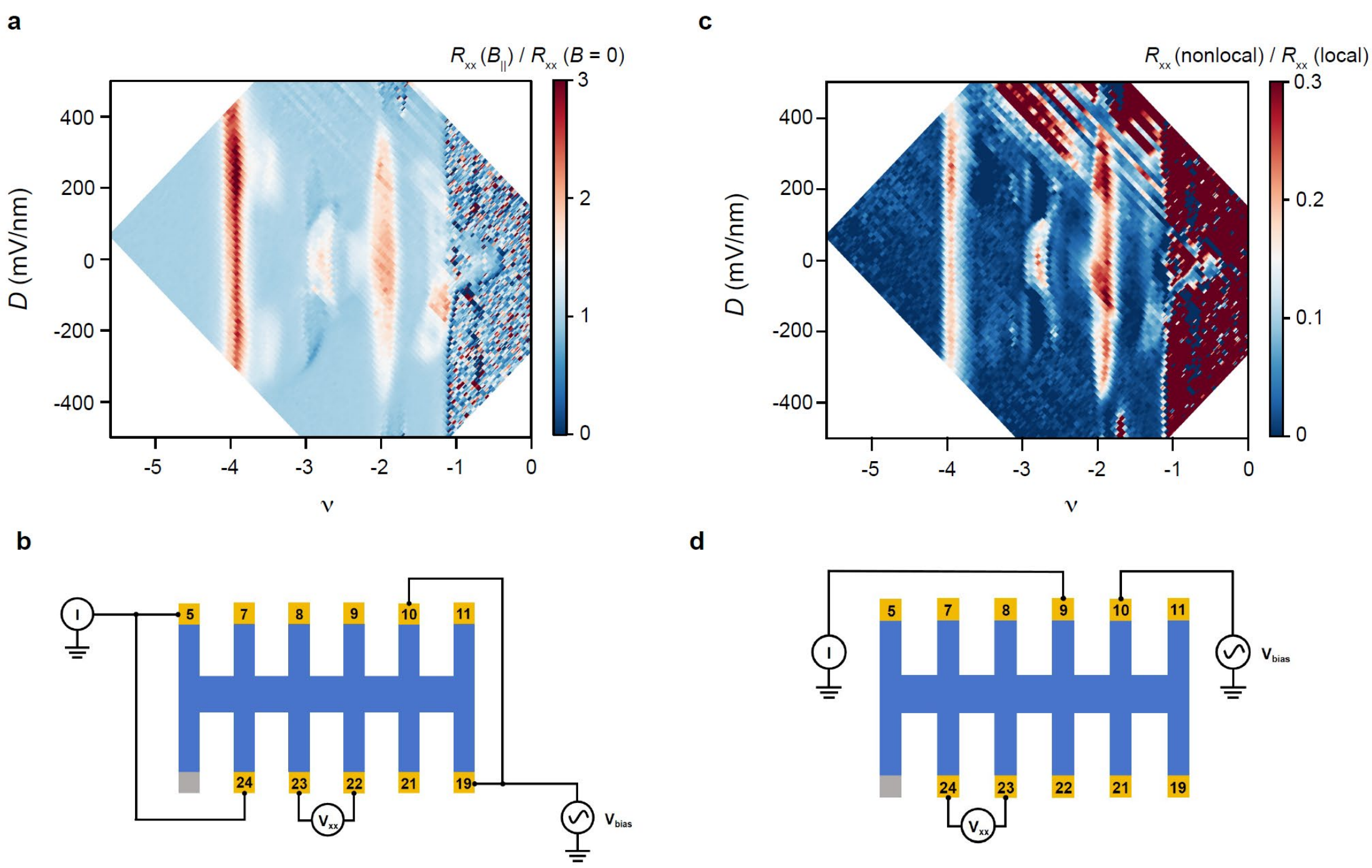

**Extended Data Fig. 7. In-plane Magnetoresistance and nonlocal transport of device B (3.0°). a**, The in-plane magnetoresistance $R_{xx}(B_{\|}=0.3\ \mathrm{T})/R_{xx}(B=0)$ versus $\nu$ and $D$ measured at $T$ = 500 mK. **c**, The ratio between nonlocal resistance and local resistance as a function of $\nu$ and $D$ measured at $T$ = 500 mK and zero magnetic field. **b** and **d** illustrate the measurement configurations in **a** and **c**, respectively.

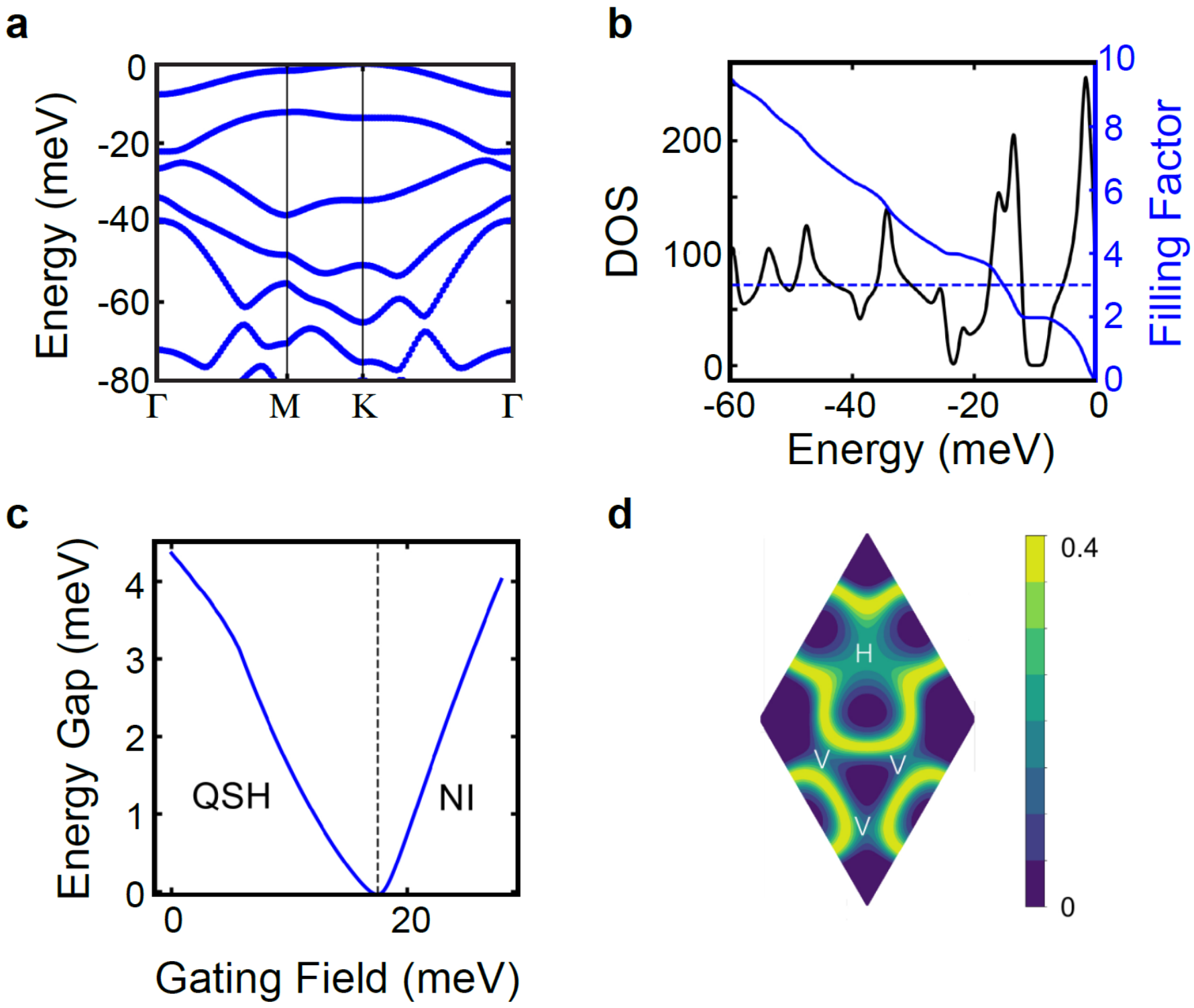

**Extended Data Fig. 8. Continuum model electronic structures at twist angle 3.15°.** (**a**) Band structure, (**b**) density of states, (**c**) gating field dependent band gap between first two moiré bands, (**d**) Fermi surface contour at $\nu$ =-3, which shows three ordinary van Hove singularity (V) and one high order van Hove singularity (H).

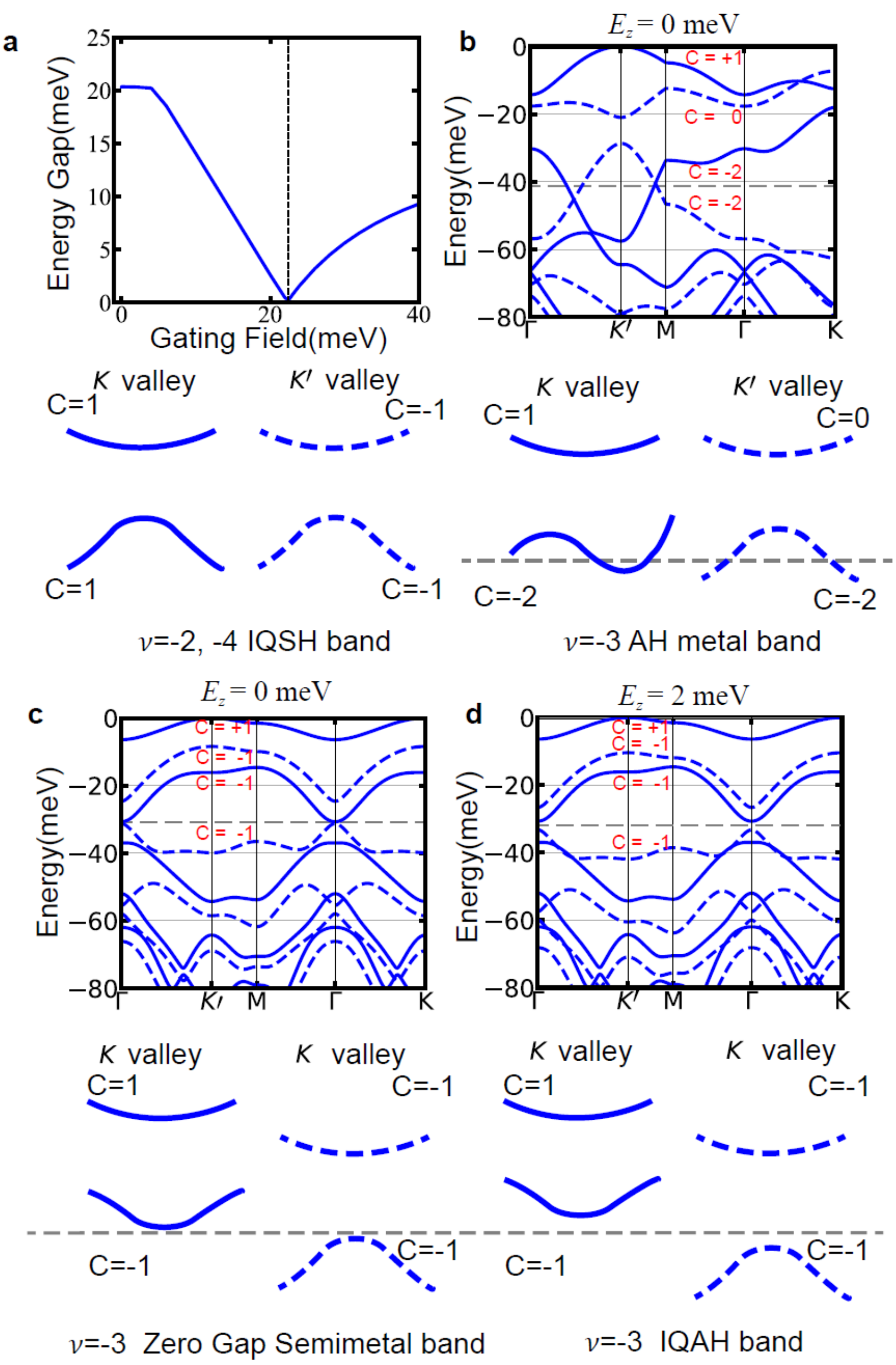

**Extended Data Fig. 9. Hartree-Fock simulation and schematic band structures of integer topological phases. a**, Hartree-Fock phase diagram at $\nu$ = -2 and $\epsilon = 30$; **b**, Hartree-Fock band structure at $\nu$ = -3 and $\epsilon = 20$ without magnetic field with non-polarized initial state, showing a partially polarized anomalous Hall metal; **c**, Hartree-Fock band structure with the same configuration but with $\epsilon = 30$, showing a nearly gapless IQAH state. **d**, Hartree-Fock band structure at and $\epsilon = 30$ with Zeeman energy $E_Z$= 2 meV, displaying a $C$ = -1 Chern insulator from two occupied K valley bands with $C$ = (1,-1), and one occupied K' valley band with $C$ = -1.

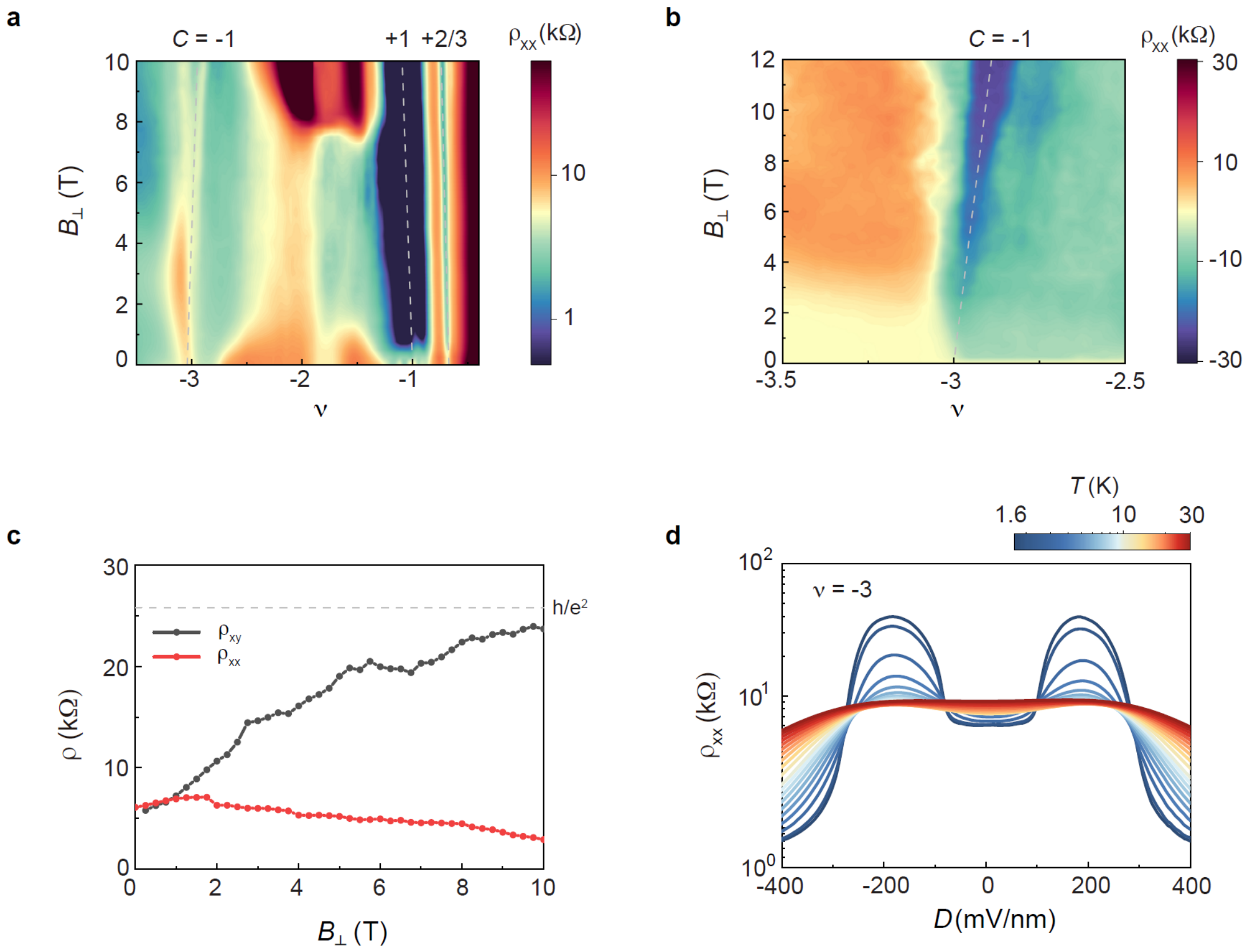

**Extended Data Fig. 10. Chern state at $\nu$ = -3 of device B (3.0°). a**,**b**, $\rho_{xx}$ (**a**) and $\rho_{xy}$ (**b**) versus $\nu$ and $B_\perp$ at $T$ = 1.6 K and $D \approx 0$ V/nm. Dashed lines represent the expected dispersions based on Streda formula for the IQAH state at $\nu$ = -1 with $C$ = 1, the FQAH state at $\nu$ = -2/3 with $C$ = 2/3, and the Chern state emerged under $B_\perp$ at $\nu$ = -3 with $C$ = -1, respectively. **c**, $\rho_{xx}$ and $\rho_{xy}$ versus $B_\perp$ alone the Streda dispersion line of the Chern state at $\nu$ = -3. The $\rho_{xy}$ approaches the expected quantized value of $h/e^2$, and the $\rho_{xx}$ is gradually vanishing with increasing $B_\perp$. **d**, Temperature dependence of $\rho_{xx}$ versus $D$ at $\nu$ = -3 from $T$ = 1.6 K to 30 K.

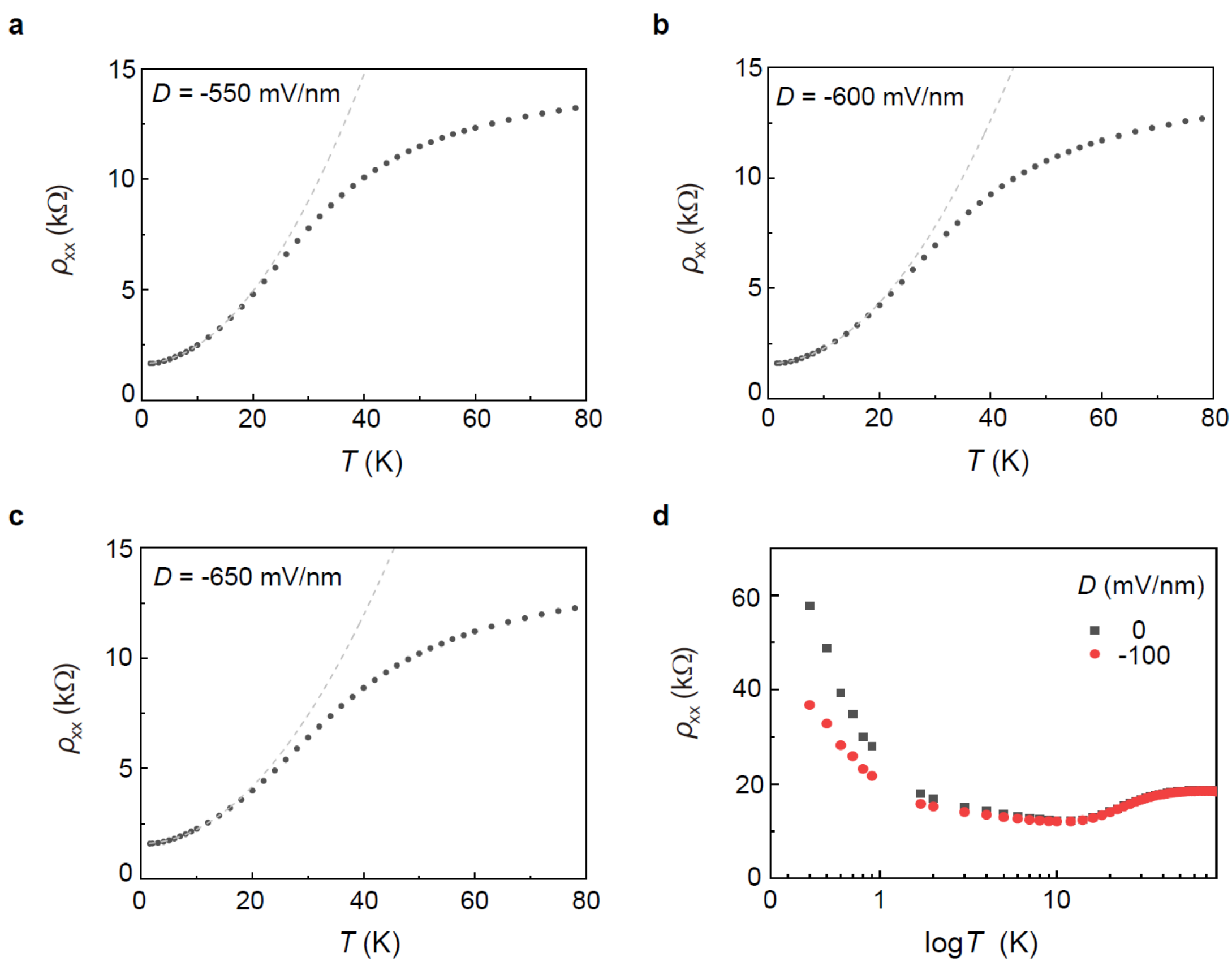

**Extended Data Fig. 11. $\rho_{xx}$ at various *D*-field at $\nu \approx -5/2$ of device A (3.15°). a-c,** Temperature dependent $\rho_{xx}$ at $D = -550$ mV/nm (**a**), -600 mV/nm (**b**), -650 mV/nm (**c**), respectively. The Fermi liquid fitting of low temperature data points ($T \approx 1.5$ K - 20 K) is shown by dashed lines in **a**-**c**. **d**, Temperature dependent $\rho_{xx}$ at $D = 0$ and -100 mV/nm, plotted on a logarithmic scale with respect to $T$. The $\rho_{xx}$ data points cannot be well fitted by a log($T$) dependence as conventional Kondo effect within a reasonable temperature range.

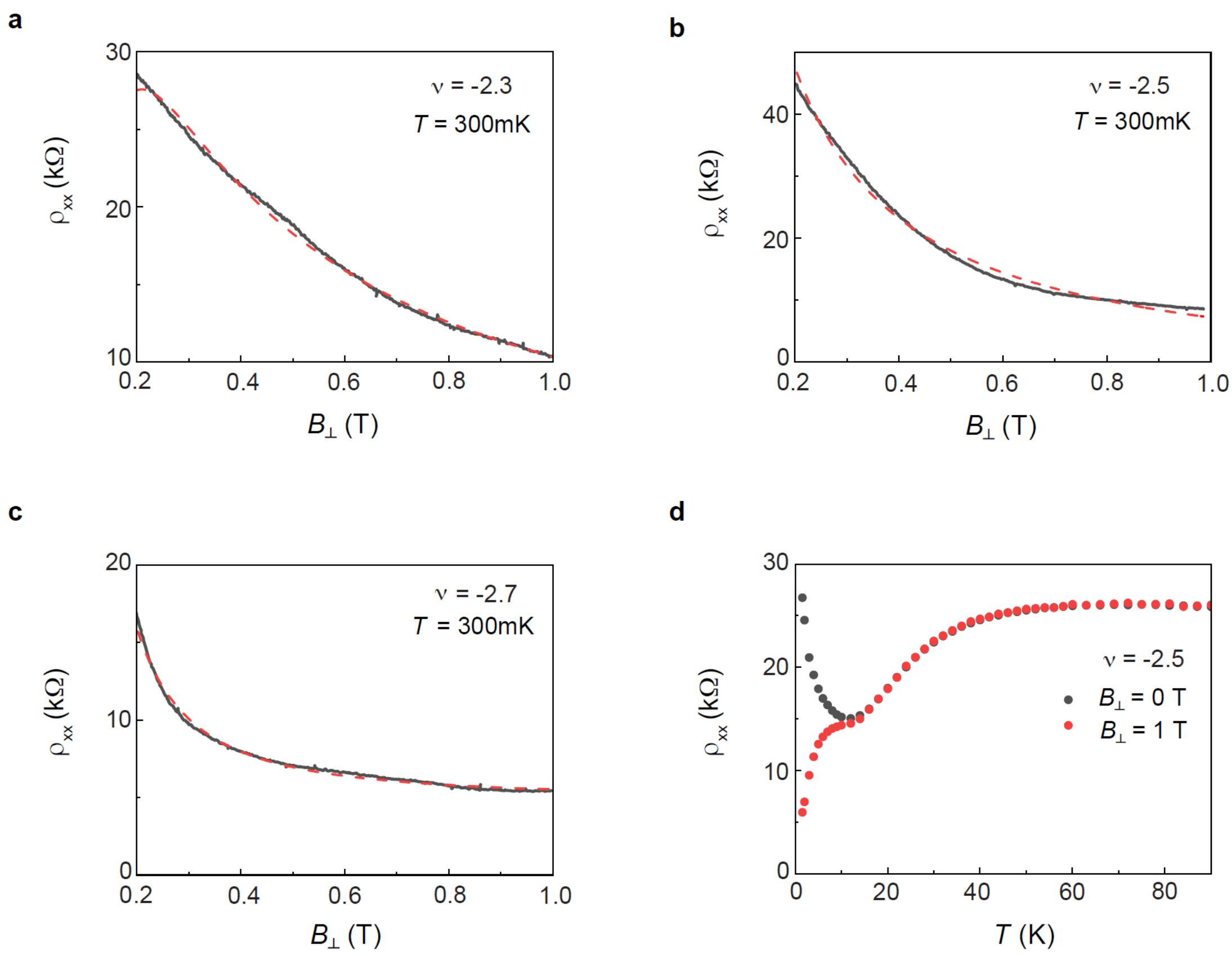

**Extended Data Fig. 12. Out-of-plane magnetoresistance fitting, *T* dependence of $\rho_{xx}$ under $B_\perp$ at $\nu \approx$ -5/2 of device A (3.15°). a-c,** $\rho_{xx}$ versus $B_\perp$ at $\nu$ = -2.3 (**a**), $\nu$ = -2.5 (**b**), $\nu$ = -2.7 (**c**). Dashed lines in **a-c** are the fitting curves of magnetoresistance with $\rho_{xx}(B_\perp) = a/B_\perp - b/B_\perp^2 + c$. The measurements in **a-c** were performed at $T$ = 300 mK and $D \approx 0$ V/nm. **d**, Temperature dependent $\rho_{xx}$ at $\nu \approx$ -2.5 and $D \approx 0$ V/nm under $B_\perp$ = 0 T and 1 T, respectively.